\documentclass[fleqn,usenatbib]{mnras}

\usepackage[utf8]{inputenc}
\usepackage{CJKutf8}
\usepackage[T1]{fontenc}

\DeclareRobustCommand{\VAN}[3]{#2}
\let\VANthebibliography\thebibliography
\def\thebibliography{\DeclareRobustCommand{\VAN}[3]{##3}\VANthebibliography}

\usepackage{graphicx,subfigure}	
\usepackage{amsmath}	
\usepackage{amssymb}	
\usepackage[english]{babel}
\usepackage[utf8]{inputenc}
\usepackage[T1]{fontenc}
\usepackage{pdflscape}	

\newcommand{\SiII}{\ion{Si}{ii}}
\newcommand{\dmvalue}{$1.04 \pm 0.04$}
\newcommand{\sbvalue}{$ 0.98 \pm 0.01 $}
\newcommand{\distmd}{$32.58 \pm 0.45$}
\newcommand{\mbmag}{$-19.52 \pm 0.47$}
\newcommand{\mbobs}{$13.55 \pm 0.02$}
\newcommand{\ebvalue}{$0.10 \pm 0.04$}
\newcommand{\tbmax}{$58510.2 \pm 0.8$}
\newcommand{\tzero}{$58494.09 \pm 0.03$}
\newcommand{\tlcv}{$11.19 \pm 0.01$}
\newcommand{\tgama}{$37.42 \pm 0.70$}
\newcommand{\trise}{$15.1 \pm 0.8$}
\newcommand{\Lmax}{$1.70 \pm 0.63\times 10^{43}$\,erg\,s$^{-1}$}
\newcommand{\MniValue}{$0.66 \pm 0.05\,M_{\odot}$}
\newcommand{\VSiII}{$10,200 \pm 141$\,km~s$^{-1}$}
\newcommand{\VSiIIDot}{$21 \pm 5$\,km\,s$^{-1}$\,d$^{-1}$}
\newcommand{\RSiII}{$0.17 \pm 0.01$}
\newcommand{\excess}{5.0\% }

\title[SN\,2019np]{Observations of the Very Young Type Ia Supernova 2019np with Early-excess Emission}

\author[Hanna Sai et al.]{
Hanna Sai,$^{1}$ \thanks{E-mail: shn17@mails.tsinghua.edu.cn}
Xiaofeng Wang,$^{1,2}$\thanks{E-mail: wang\_xf@mail.tsinghua.edu.cn}
Nancy Elias-Rosa$^{3,4}$,
Yi Yang\begin{CJK*}{UTF8}{gbsn}
(杨轶)
\end{CJK*}$^{5,\mathsection}$
Jujia Zhang,$^{6,7,8}$
\newauthor
Weili Lin,$^{1}$
Jun Mo,$^{1}$
Anthony L. Piro,$^{9}$
Xiangyun Zeng,$^{10}$
Reguitti Andrea,$^{11,12,3}$ 
\newauthor
Peter Brown,$^{13}$
Christopher R. Burns,$^{9}$
Yongzhi Cai,$^{1}$ 
Achille Fiore,$^{3}$
Eric Y. Hsiao,$^{14}$
\newauthor
Jordi Isern,$^{14,15}$
K. Itagaki,$^{16}$
Wenxiong Li,$^{17}$
Zhitong Li,$^{18,19}$
Priscila J. Pessi,$^{20}$
\newauthor
M. M. Phillips,$^{21}$
Stefan Schuldt,$^{22,23}$
Melissa Shahbandeh,$^{14}$
Maximilian D. Stritzinger,$^{24}$
\newauthor
Lina Tomasella,$^{3}$
Christian Vogl,$^{22}$
Bo Wang,$^{6}$
Lingzhi Wang,$^{25,26}$
Chengyuan Wu$^{6,1}$
\newauthor
Sheng Yang,$^{27}$
Jicheng Zhang,$^{28}$
Tianmeng Zhang,$^{18,19}$
Xinghan Zhang,$^{1}$\\
$^{1}$Physics Department and Tsinghua Center for Astrophysics (THCA), Tsinghua University, Beijing, 100084, China\\
$^{2}$Beijing Planetarium, Beijing Academy of Science and Technology, Beijing 100044, China\\
$^{3}$INAF – Osservatorio Astronomico di Padova, Vicolo dell'Osservatorio 5, 35122 Padova, Italy\\
$^{4}$Institute of Space Sciences (ICE, CSIC), Campus UAB, Carrer de Can Magrans s/n, 08193 Barcelona, Spain\\ 
$^{5}$Department of Astronomy, University of California, Berkeley, CA 94720-3411, USA \\
$^{\mathsection}$Bengier-Winslow-Robertson Fellow \\
$^{6}$Yunnan Observatories (YNAO), Chinese Academy of Sciences, Kunming 650216, China\\
$^{7}$Key Laboratory for the Structure and Evolution of Celestial Objects, Chinese Academy of Sciences, Kunming 650216, China\\
$^{8}$Center for Astronomical Mega-Science, Chinese Academy of Sciences, 20A Datun Road, Chaoyang District, Beijing, 100012, China\\
$^{9}$Carnegie Observatories, 813 Santa Barbara Street, Pasadena, CA 91101, USA\\
$^{10}$Center for Astronomy and Space Sciences, China Three Gorges University, Yichang, 443000, China\\
$^{11}$Departamento de Ciencias F\'{i}sicas – Universidad Andrés Bello, Avda. Rep\'{u}blica 252, Santiago 8320000, Chile\\
The remaining affiliations can be found after the references.\\}
\date{}

\pubyear{2022}

\begin{document}
\label{firstpage}
\pagerange{\pageref{firstpage}--\pageref{lastpage}}
\maketitle

\begin{abstract}
Early-time radiative signals from type Ia supernovae (SNe\,Ia) can provide important constraints on the explosion mechanism and the progenitor system. We present observations and analysis of SN\,2019np, a nearby SN\,Ia discovered within 1-2 days after the explosion. Follow-up observations were conducted in optical, ultraviolet, and near-infrared bands, covering the phases from $\sim-$16.7 days to $\sim$+367.8 days relative to its $B-$band peak luminosity. The photometric and spectral evolutions of SN\,2019np resembles the average behavior of normal SNe\,Ia. The absolute $B$-band peak magnitude and the post-peak decline rate are $M_{\rm max}(B)=$\mbmag\,mag and $\Delta m_{\rm15}(B)$ =\dmvalue\,mag, respectively. No Hydrogen line has been detected in the near-infrared and nebular-phase spectra of SN\,2019np. Assuming that the $^{56}$Ni powering the light curve is centrally located, we find that the bolometric light curve of SN\,2019np shows a flux excess up to \excess in the early phase compared to the radiative diffusion model. Such an extra radiation perhaps suggests the presence of an additional energy source beyond the radioactive decay of central nickel. Comparing the observed color evolution with that predicted by different models such as interactions of SN ejecta with circumstellar matter (CSM)/companion star, a double-detonation explosion from a sub-Chandrasekhar mass white dwarf (WD), and surface $^{56}$Ni mixing, the latter one is favored.
\end{abstract}
\begin{keywords}
supernovae: general – supernovae: individual: (SN\,2019np)
\end{keywords}


\section{Introduction}
Since the late twentieth century, the studies of Type Ia supernovae (SNe\,Ia) have led to the discovery of the accelerating universe \citep{1998AJ....116.1009R, 1999ApJ...517..565P,2016ApJ...826...56R,2018ApJ...853..126R,2021arXiv211204510R}. Nowadays, SNe\,Ia are widely used as standardizable candles in observational cosmology \citep{1996ApJ...473...88R,2005ApJ...620L..87W,2005A&A...443..781G,2006Natur.443..308H,2011NatCo...2..350H,2018ApJ...869...56B}. However, the exact formation channel and explosion physics of SNe\,Ia are still unknown \citep{2012NewAR..56..122W,2013Sci...340..170W,2014ARA&A..52..107M,2019NatAs...3..706J}.

Three popular scenarios for the explosion of SNe\,Ia are: 
1) The `single-degenerate' (SD) channel, in which a carbon–oxygen (CO) WD accretes matter from a nondegenerate companion such as a main-sequence star, a giant, or a He star. A thermonuclear explosion process is expected to occur when the progenitor CO WD reaches the critical Chandrasekhar mass of $M_{\rm Ch}\sim1.4 M_{\odot}$ \citep{1973ApJ...186.1007W,1997Sci...276.1378N,2008NewAR..52..381P,2009MNRAS.395..847W}; 
2) The `double-degenerate' (DD) channel, in which the explosions of SNe\,Ia are triggered by the compressional heat generated by the dynamical merging process of two CO WDs \citep{1984ApJ...277..355W, 1984ApJS...54..335I, 1990ApJ...348..647B, 2012ApJ...747L..10P}; 
3) The direct head-on collision of two WDs in triple systems \citep{2013ApJ...778L..37K}; however, the rate of head-on collisions of WDs in triple systems has been suggested too low to account for the majority of SNe\,Ia explosions \citep{2018A&A...610A..22T}. In the SD channel, an extended CSM prorfile is expected to form around the progenitor system during its evolution towards a SN\,Ia explosion. Although the signature of CSM has been revealed for some SNe\,Ia via high-resolution spectroscopy \citep{2003Natur.424..651H,2006ApJ...650..510A,2007Sci...317..924P,2011Sci...333..856S,2012Sci...337..942D,2013MNRAS.436..222M,2013ApJS..207....3S}, the lack of narrow H emission lines in the late-time spectra still challenges the SD model \citep{2005A&A...443..649M,2007ApJ...670.1275L,2013ApJ...762L...5S,2016MNRAS.457.3254M,2020MNRAS.493.1044T}. Additionally, searches for pre-explosion or surviving companions of the progenitor through deep images \citep{2011Natur.480..348L,2012Natur.489..533G,2012Natur.481..164S} have firmly excluded the presence of giant and subgiant companions. 

On the other hand, detection of the first light after a SN explosion can also provide important constraints on progenitors of SNe\,Ia \citep{2010ApJ...708.1025K}. In a SD progenitor system, the SN ejecta could run into the nondegenerate companion or surrounding CSM. Then an excess of ultraviolet (UV)/optical emission could be generated and become detectable within the first few days for the interaction between SN ejecta and a companion star or within a few hours for the interaction between SN ejecta and ambient CSM. For the former case, the flux excess can last for a few days, depending on the size of the companion, the progenitor-companion separation, the ejecta velocity, and the viewing angle. During the \emph{Kepler} mission \citep{2010ApJ...713L.115H}, a few SNe were detected in the \emph{Kepler} fields, including two spectroscopically confirmed SNe\,Ia and three possible ones. There is no signature of ejecta interaction with a stellar companion for the latter three (KSN 2012a, KSN 2011b and KSN 2011c) \citep{2015Natur.521..332O} and SN\,2018agk \citep{2021ApJ...923..167W}. Blue excess flux was detected in the early-time light curve of SN\,2018oh, however, the explanations are not converged \citep{2019ApJ...870L...1D,2019ApJ...870...13S,2019ApJ...870...12L}. For example, both the deep mixing carbon feature and nondetection of hydrogen in the nebular-phase spectrum of SN\,2018oh do not favour a SD scenario for its origin \citep{2019ApJ...870...12L,2019ApJ...872L..22T}. A similar ``blue bump'' feature was also detected in the nearby SN\,2017cbv \citep{2017ApJ...845L..11H}, but analysis of the nebular-phase spectroscopy argues against a nondegenerate H- or He-rich companion as its progenitor \citep{2018ApJ...863...24S}. For the latest case, however, the extra emission in the early time usually lasts for a few hours, depending on the density and distance of the CSM. In a recent study, SN\,2020hvf is reported to have such an early-time excess emission \citep{2021ApJ...923L...8J}. 

Besides interaction with a companion star or CSM disk, the early excess emission may also be produced by other mechanisms, e.g., radioactive decay of $^{56}$Ni near the surface of the exploding WD \citep{2013ApJ...769...67P,2020A&A...634A..37M}. Mixing of $^{56}$Ni into the outer layers would result in bluer and steeper rising light curves \citep{2016ApJ...826...96P} and this mechanism works for both SD and DD progenitors. \cite{2020A&A...634A..37M} shows that the observed diversity in light curves of SNe\,Ia can be reproduced by varying the distribution of $^{56}$Ni. Another explosion model that can produce such $^{56}$Ni distribution  is the double-detonation explosion of a sub-Chandrasekhar mass WD, where the initial detonation of the outer-layer He would produce a certain amount of nickel near the surface and it sends a shockwave to the C/O core to trigger the second detonation \citep{1994ApJ...423..371W,2013A&A...559A..94W,2017MNRAS.472.2787N}, thus leading to a wide range of absolute magnitudes and colors. However, in order to match the observed peak brightness of SNe\,Ia, this model requires that the number of iron-group elements (IGEs) synthesized during He shell detonation is very small. The radioactive decay of the surface material would produce photons and they diffuse out of the ejecta rapidly, which results in a flux excess in the first few days after explosion.
 
The detection of the very young thermonuclear SN\,2019np provides us a rare opportunity to examine the first-light evolution of a SN\,Ia. In this paper, we present extensive follow-up observations of SN\,2019np within the UV to NIR wavelength ranges. We also analyze its photometric and spectroscopic behaviours and compare the observational properties of SN\,2019np with those of other well-studied SNe\,Ia. Observations and data reduction are outlined in Section~\ref{sec:obser}, the light/color curves are described in Section~\ref{sec:lc}, and the spectroscopic behavior of SN\,2019np and its temporal evolution are presented in Section~\ref{sec:spec}. We discuss the properties of SN\,2019np and its explosion parameters in Section~\ref{sec:dis}. The conclusions are given in Section~\ref{sec:con}.

\section{observations} \label{sec:obser}
\begin{figure}
    \centering
	\includegraphics[width=0.9\columnwidth]{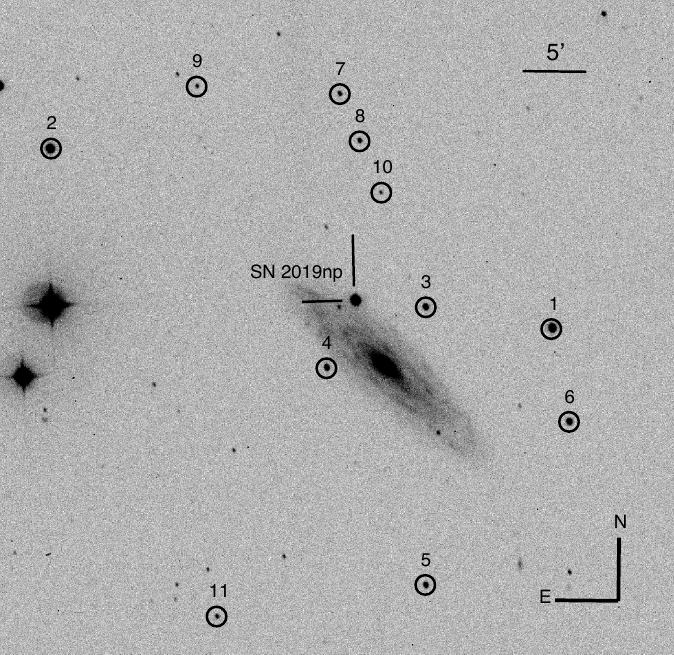}
    \caption{Image of SN\,2019np in NGC 3254, taken with Tsinghua-NAOC 0.8~m telescope. Reference stars are marked with black circles and listed in Table \ref{standstar}. North is up and east is left.}
\label{verify}
\end{figure}

SN\,2019np was discovered on 2019 Jan. 9.67 (UT) by K. Itagaki in NGC 3254 which is an Sbc-type spiral galaxy (see Figure~\ref{verify}) with a redshift of only 0.00452. The classification spectrum obtained $\sim$one day after the discovery suggests that SN\,2019np was a very young, normal SN\,Ia, at $\sim$15 days before the maximum luminosity \citep{2019ATel12374....1W,2019ATel12375....1K}. 

\subsection{Photometry}
Photometric observations were obtained through various facilities including: (1) the 0.8~m Tsinghua-NAOC Telescope (TNT) in China \citep{2012RAA....12.1585H}; (2) the Lijiang 2.4~m Telescope (LJT) at Yunnan Observatory; (3) the 0.8~m Joan Or\'o Telescope+MEIA3 at Montsec Astronomical Observatory (Spain); (4) the 0.67/0.92 m Schmidt Telescope at Mount Ekar Observatory (Italy); (5) the 1.8~m Copernico Telescope+AFOSC at Mount Ekar Observatory (Italy); (6) the 2.6~m Nordic Optical Telescope (NOT)+ALFOSC at the Roque de Los Muchachos Observatory (Spain); (7) the 0.6~m reflector telescope+BITRAN-CCD at Itagaki Astronomical Observatory (Japan). All images were pre-processed, including bias subtraction, flat-field correction, and cosmic-ray rejection with the standard IRAF\footnote{{IRAF} is distributed by the National Optical Astronomy Observatories, which are operated by the Association of Universities for Research in Astronomy, Inc., under cooperative agreement with the National Science Foundation (NSF).} packages. We use the pipeline \texttt{Zuruphot} developed for automatic photometry on TNT (Mo et al. in prep) to perform point-spread-function (PSF) photometry for both SN\,2019np and local reference stars. 

The instrumental magnitudes were then converted to the Johnsons $BV$ \citep{1966CoLPL...4...99J} and Sloan Digital Sky Survey (SDSS) $gri$-band photometry \citep{1996AJ....111.1748F}, based on the magnitude of 10 relatively bright local comparison stars from the SDSS Data Release 9 catalog. SN\,2019np was also observed by the Ultra-violet/Optical Telescope (UVOT; \citealp{2004ApJ...611.1005G,2005SSRv..120...95R}) on the Neil Gehrels Swift Observatory \citep{2004ApJ...611.1005G} in three UV ($uvw2$, $uvm2$, $uvw1$) and three optical filters ($u$, $b$, $v$). Photometry was extracted using the HEASOFT \footnote{HEASOFT, the High Energy Astrophysics Software, \url{https://www.swift.ac.uk/analysis/software.php}} with the latest \emph{Swift} calibration database \footnote{\url{https://heasarc.gsfc.nasa.gov/docs/heasarc/caldb/swift/}}. The final Optical and UV magnitudes are reported in Table~\ref{tab:optical} and \ref{swift}, respectively.

\subsection{Spectroscopy}
Optical spectra of SN\,2019np were obtained with the Lijiang 2.4~m telescope (LJT+YFOSC); Xinglong 2.16~m telescope (XLT+BFOSC) \citep{2016PASP..128j5004Z}; the 1.8~m Copernico Telescope+AFOSC at the Mount Ekar Observatory (Italy); the Alhambra Faint Object Spectrograph Camera (ALFOSC) on the 2.56~m Nordic Optical Telescope (NOT\footnote{\url{http://www.not.iac.es/instruments/alfosc/}}); the 10.4~m Gran Telescopio CANARIAS (GTC)+OSIRIS at the Roque de Los Muchachos Observatory (Spain); and the 2.0~m Faulkes Telescope North (FTN)+FLOYDS at Haleakala Observatory (USA), spanning from $-16.7$ to 367.8 days (d) relative to the $B-$band maximum light. A journal of spectroscopic observations of SN\,2019np is presented in Table~\ref{log}. We reduced all the spectra using the standard IRAF routine. Flux calibration of the spectra was performed with spectrophotometric standard stars taken on the same nights. We correct the atmospheric extinction using the extinction curves of local observatories. The telluric correction was derived from the spectrophotometric standard star spectral observations and applied to the SN spectra.

Two near-infrared (NIR) spectra of SN\,2019np were obtained using the Folded port InfraRed Echellette (FIRE, \citealp{2013PASP..125..270S}) spectrograph mounted on the 6.5-m Magellan Baade telescope at Las Campanas Observatory, Chile. The spectra cover a wavelength range of 0.8-2.5 $\mu m$ with the high-throughput prism mode coupled to a 0."6 slit. For telluric correction, we observed an A0V star close in time and at approximately similar air mass to SN\,2019np \citep{2015A&A...578A...9H,2019PASP..131a4002H}. The spectra were reduced using the IDL pipeline \texttt{firehose} \citep{2013PASP..125..270S}. We also present two NIR spectra obtained with the SpeX spectrograph \citep{2003PASP..115..362R} mounted on the NASA Infrared Telescope Facility (IRTF) in Hawaii. A short cross-dispersed mode has been used and a 0."5 slit was placed on the target, providing a wavelength coverage of 0.8-2.5 $\mu m$. The SpeX data were reduced using the IDL code \texttt{Spextool} \citep{2004PASP..116..362C}. A journal of spectroscopic observations of SN\,2019np is presented in Table~\ref{lognir}.

\section{Photometry}\label{sec:lc}
\subsection{Optical and UV Light Curves}
\begin{figure*}
\centering
	\includegraphics[width=1.2\columnwidth]{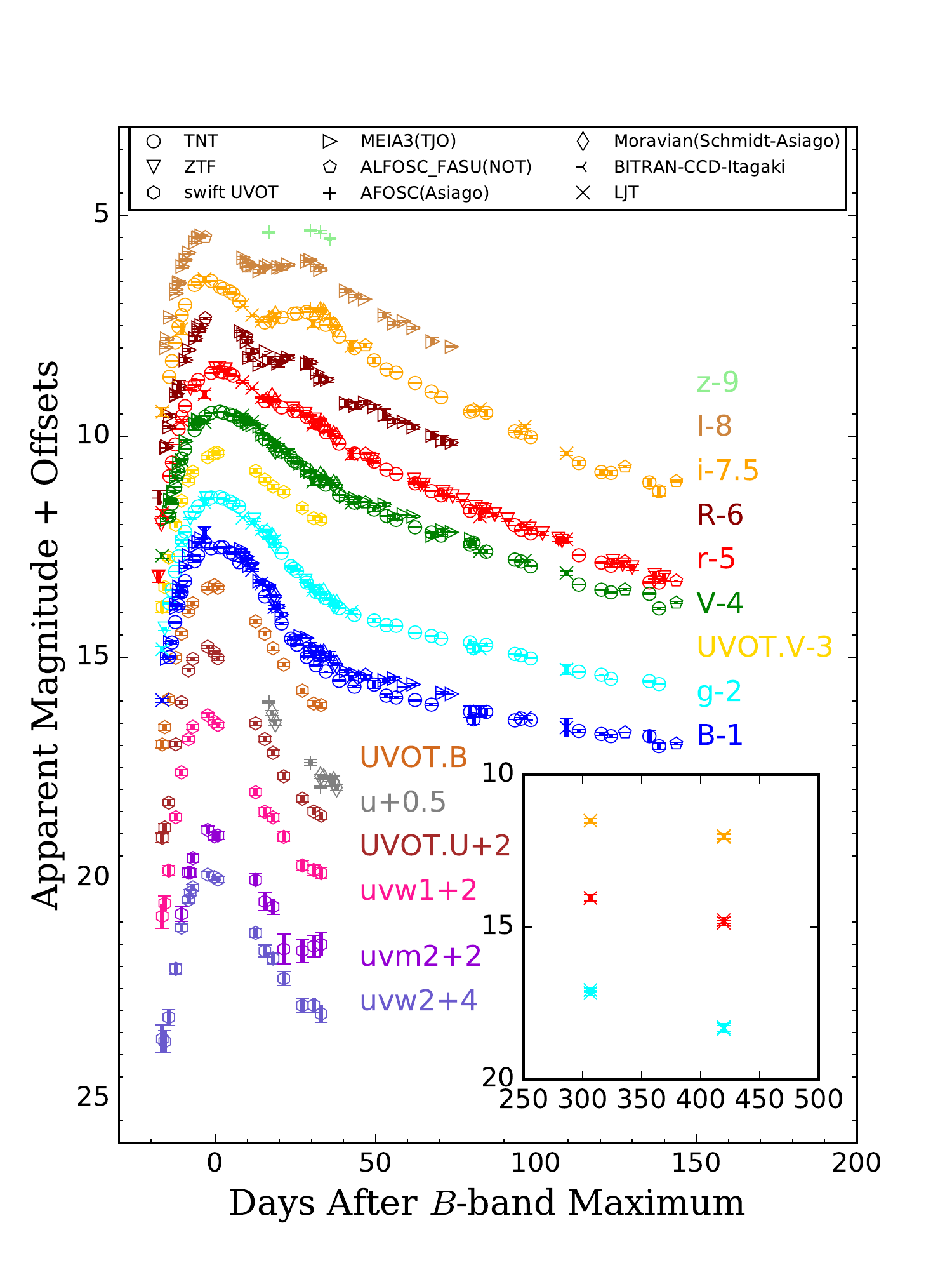}
    \caption{Optical and UV light curves of SN\,2019np.
Different colors represent different bands, including the \emph{Swift uvw2, uvm2, uvw1,} UVOT \emph{U, u}, UVOT \emph{B, B, g,} UVOT \emph{V, V, r, R, i, I} and \emph{z}. The insert panel shows the late-time light curves.}
    \label{lc}  
\end{figure*}

Figure~\ref{lc} shows the UV and optical light curves of SN\,2019np sampled during the period from -16.5 to +32.8 days and that from -17.7 to 419.5 days relative to the $B-$band maximum light, respectively. 
Similar to other normal SNe\,Ia, the light curves of SN\,2019np show shoulders in the $R/r$ bands. Secondary peaks can be identified in $I/i$-band light curves, for which the first peak appears about 2 days earlier than the $B-$band peak. 

\begin{figure*}
          \centering
   \subfigure{
     \includegraphics[scale=0.43]{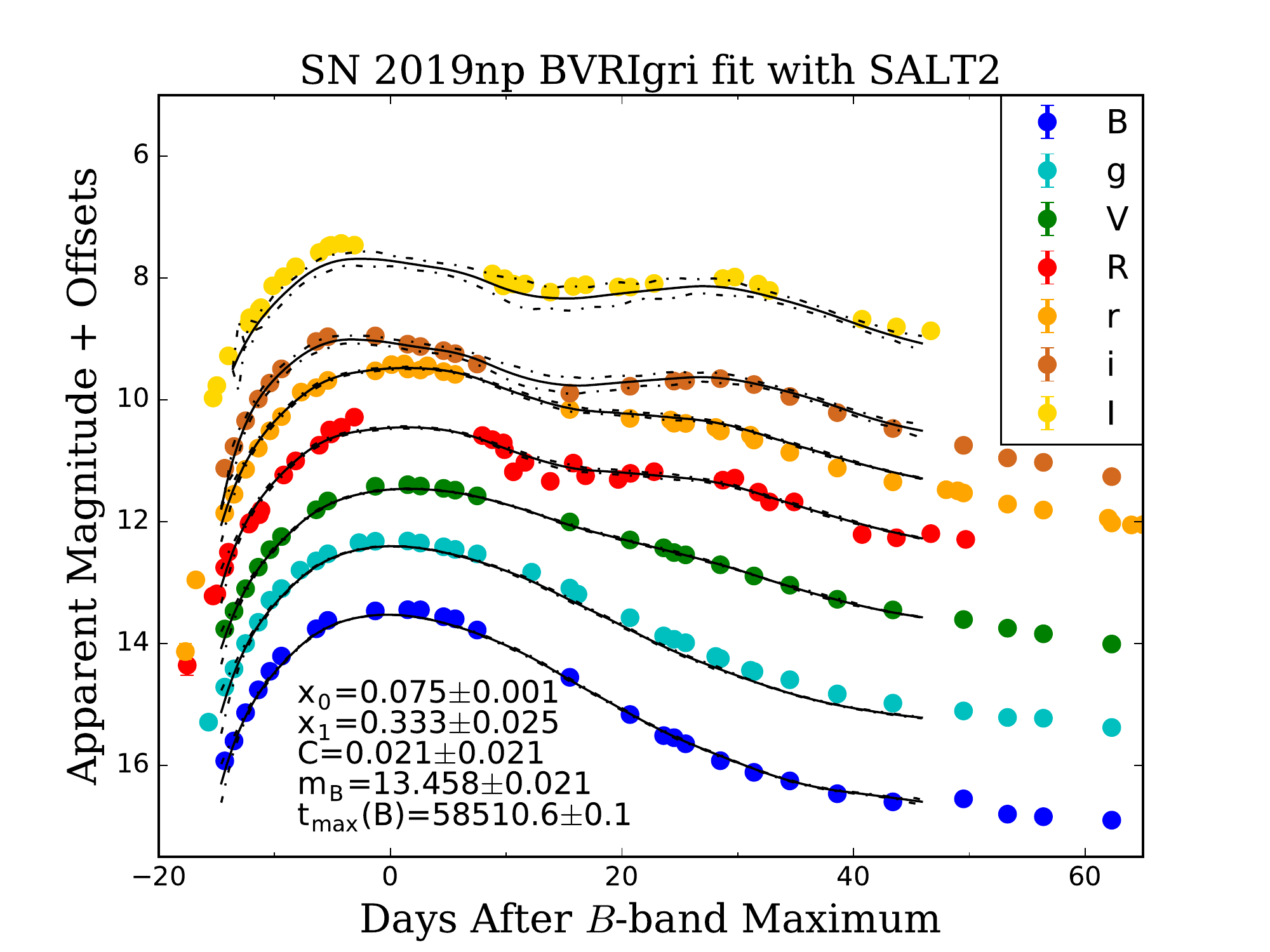}}
    \subfigure{
          \includegraphics[scale=0.43]{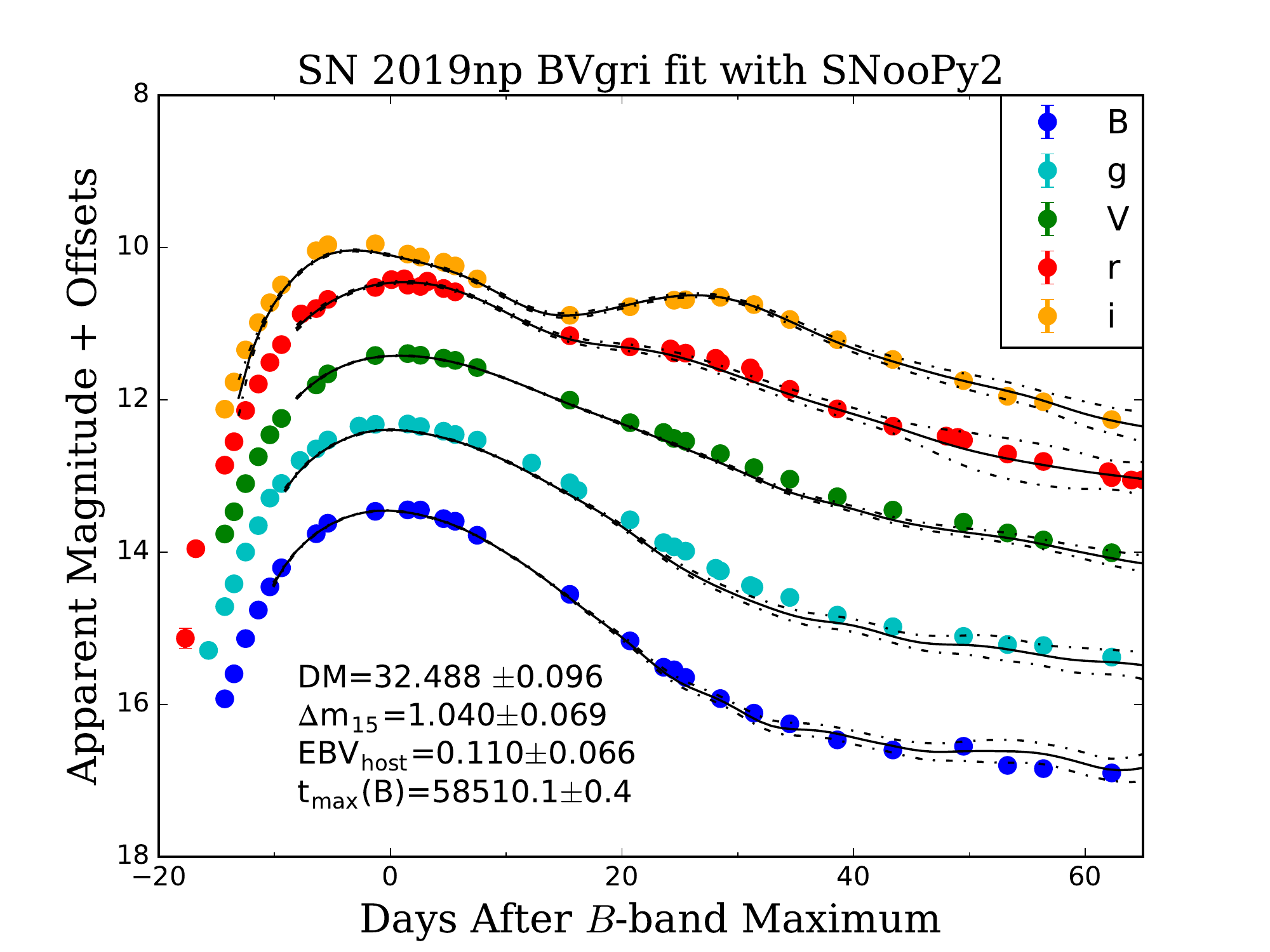}}
          \label{snoopy}
\caption{Best-fit light-curve models from SALT2 (left panel) and SNooPy2 (right panel). The light curves of different bands have been shifted vertically for better display. The dashed lines represent the 1-$\sigma$ uncertainty of the light-curve templates. The parameter $x_0$ represents the normalization of the SED sequence, $x_1$ is the light curve shape parameter and C stands for the color. $m_\textup{B}$ and $t_{\textup{max}}$(B) denote the maximum magnitude of the $B-$band and the corresponding time. DM is the distance modulus.}
\label{salt}
\end{figure*}

We applied high-order polynomial fits to the light curves of SN\,2019np around the maximum light and estimated that it reached a $B-$band peak magnitude of \mbobs\,mag on MJD \tbmax\,and a $V-$band peak magnitude of 13.62$\pm0.15$ mag on $58512.5\pm0.7$, respectively. The post-peak magnitude decline in 15 days from the $B-$band peak, $\Delta m_{\rm15}(B)$ \citep{1999AJ....118.1766P}, as determined using the light-curve fitting tools SALT2 \citep{2010A&A...523A...7G} and SuperNovae in object-oriented Python (SNooPY2, \citealp{2011AJ....141...19B}), is estimated as $1.038\pm0.004$ and $1.040\pm0.069$ mag, respectively. The best-fit light-curve models and the associated parameters are presented in Figure~\ref{salt}. 

\begin{figure}
	\includegraphics[width=\columnwidth]{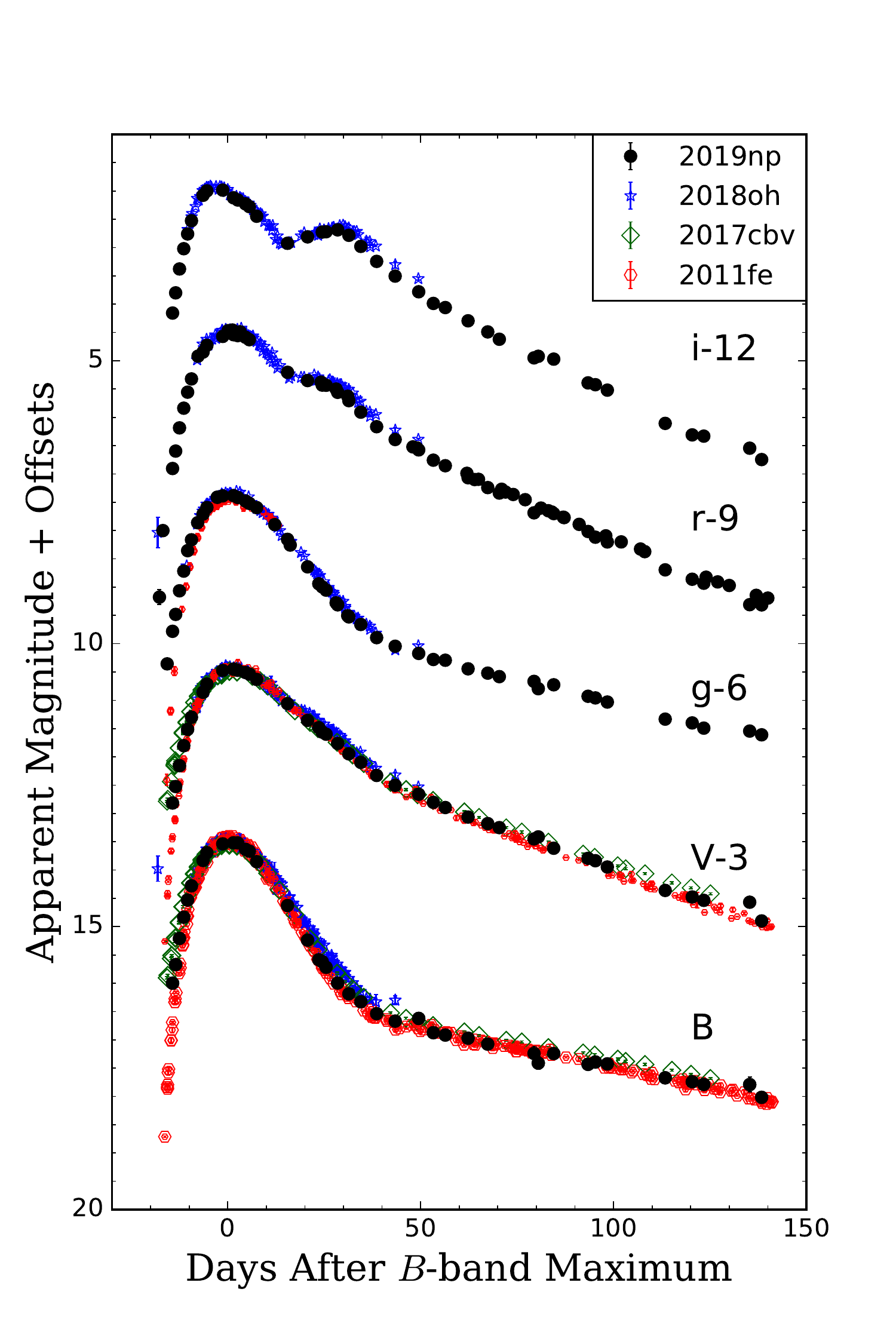}
    \caption{Comparison of the optical light curves of SN\,2019np with other well-observed SNe\,Ia with similar decline rates. The light curves of the comparison SNe\,Ia have been normalized to match the peaks of SN\,2019np.}
    \label{lccompare}
\end{figure}

\begin{figure}
	\includegraphics[width=\columnwidth]{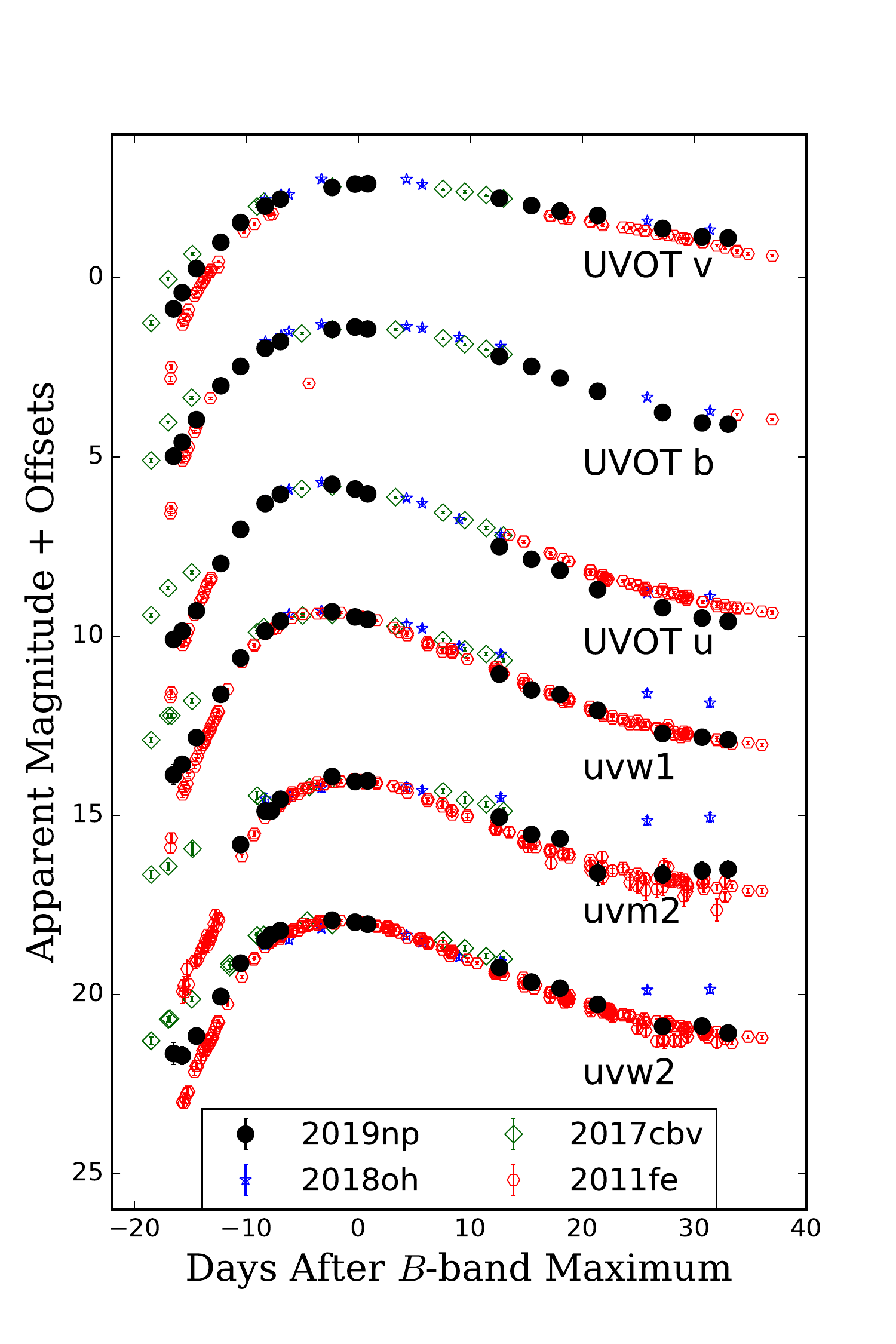}
    \caption{The UV light curves of SN\,2019np compared with other well-sampled SNe\,Ia with similar decline rates. The light curves of the comparison SNe\,Ia are normalized to match the peaks SN\,2019np.}
    \label{uvotcompare}
\end{figure}

In Figures~\ref{lccompare} and \ref{uvotcompare} we compare the optical and UV-band light curves, respectively, between SN\,2019np and several other well-observed normal SNe\,Ia with similar values of $\Delta m_{\rm15}$. The comparison sample include SN\,2018oh \citep{2019ApJ...870...12L}, SN\,2017cbv \citep{2017ApJ...845L..11H,2020ApJ...904...14W} and SN\,2011fe \citep{2012MNRAS.425.1789S,2013CoSka..43...94T,2016ApJ...820...67Z,2019MNRAS.490.3882S}. The optical light curves of SN\,2019np exhibit high similarities to those of SN\,2017cbv ($\Delta m_{\rm15}(B)= 1.06\pm0.3$ mag). Closer inspection of the UV light curves reveals that SN\,2019np shows exceptionally blue UV radiation in the very early phase, which is also similar to SN\,2017cbv. 

\subsection{Reddening}
The Galactic extinction toward SN\,2019np is estimated as $A_V ({\rm Gal}) = 0.055\,$mag according to the dust map derived by \cite{2011ApJ...737..103S}. Adopting the \cite{1989ApJ...345..245C} extinction law with a total-to-selective extinction ratio of 3.1, the reddening due to Milky Way is $E(B-V)_{\rm Gal}=0.018$ mag. This is consistent with the weak \ion{Na}{i}D absorption due to the Milky Way. 

Furthermore, we examined the absorption of the interstellar \ion{Na}{i}D doublet (5895.92, 5889.95 \AA) due to the host galaxy and derived an equivalent width (EW) of 0.68\AA~for SN\,2019np from its near-maximum-light spectra. This corresponds to a host-galaxy reddening of 0.10$\pm$0.02 mag according to the empirical relation proposed by \cite{2012MNRAS.426.1465P} (i.e., $\log_{10}(E(B-V))=1.17\times\rm{EW}(\rm{D_{1}}+\rm{D_{2}})-1.85$), and a similar result with larger dispersion has been updated by \cite{2013ApJ...779...38P}. We also employed SNooPy2 \citep{2011AJ....141...19B} to fit the multi-band light curves of SN\,2019np to determine the host-galaxy reddening, which gives $E(B-V)_{\rm host}=0.110 \pm0.066\,$mag. The best-fit results are also shown in Figure~\ref{salt}. We averaged the reddening estimated by the above methods and adopted $E(B-V)_{\rm host}$=\ebvalue\,mag as the final value. Moreover, an extinction law with $R_V=3.1$ \citep{1989ApJ...345..245C} is adopted throughout this paper. 

After correcting for the reddening due to the Milky Way and host galaxy, the $B-V$ color of SN\,2019np is found to be $-0.06 \pm 0.03$ mag around the $B-$band maximum light, in consistent with that of a normal SN\,Ia (see, e.g., \citealt{2009ApJ...697..380W}).

\subsection{Color Curves}
\begin{figure*}
	\includegraphics[width=2\columnwidth]{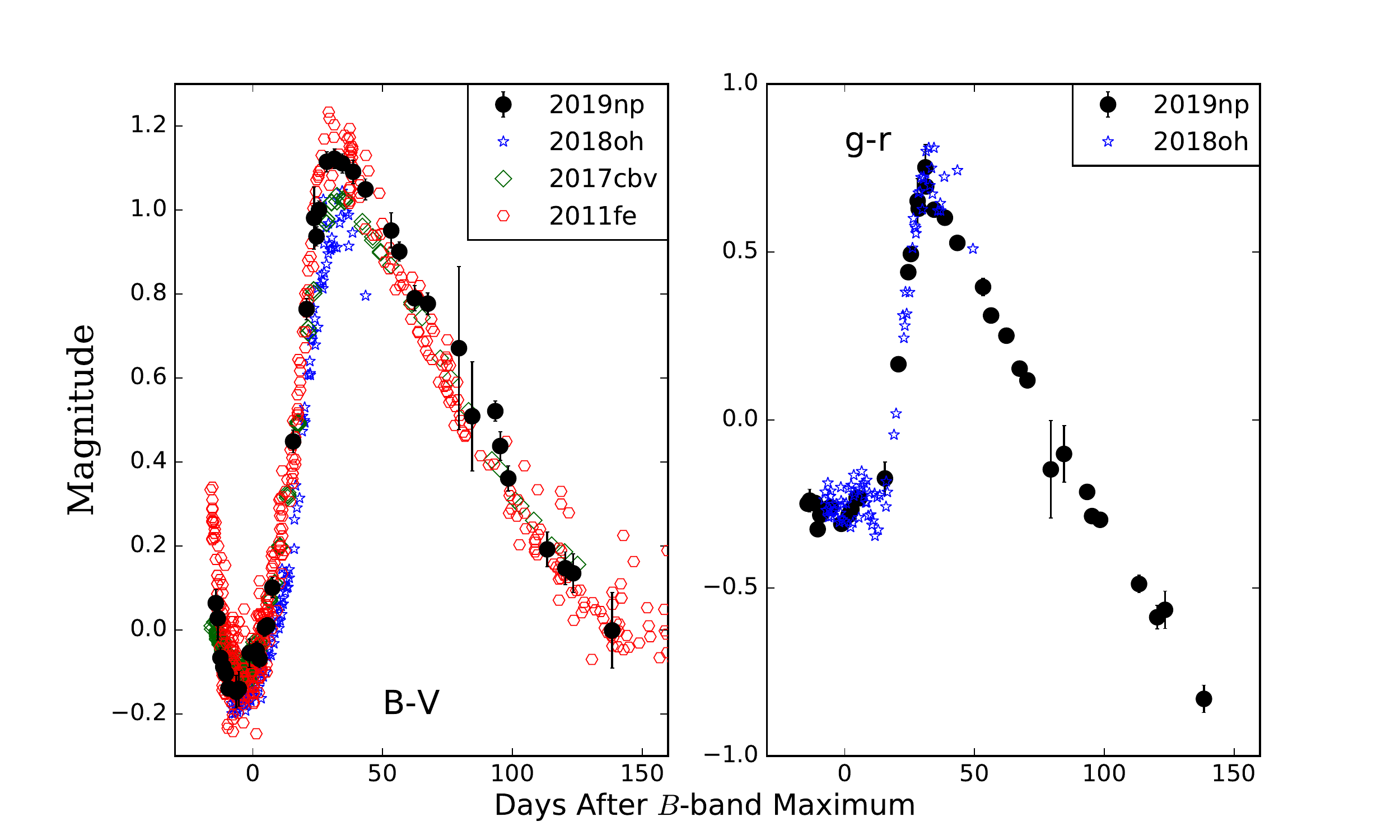}
    \caption{Reddening-corrected $B-V$ (left panel) and $g-r$ (right panel) color curves of SN\,2019np, compared with those of SNe 2018oh, 2017cbv and 2011fe.}
    \label{opticalcompare}
\end{figure*}

In Figure~\ref{opticalcompare}, we compare the color evolution of SN\,2019np with that of some well-observed SNe\,Ia with similar $\Delta m_{\rm15}(B)$. At early times, SN\,2019np showed blue colors similar to SN\,2011fe, SN\,2017cbv and SN\,2018oh. After that, the $B-V$ and $g-r$ color curves evolve bluewards and reach the minimum value at about 10 days prior to the $B$-band maximum. Then they evolve redwards again and reach the red peak at t$\sim30$ d; and they gradually become blue after the red peak. 

\section{Spectroscopy}\label{sec:spec}

\subsection{Temporal Evolution of the Optical Spectra}
\begin{figure*}
	\includegraphics[width=1.45\columnwidth]{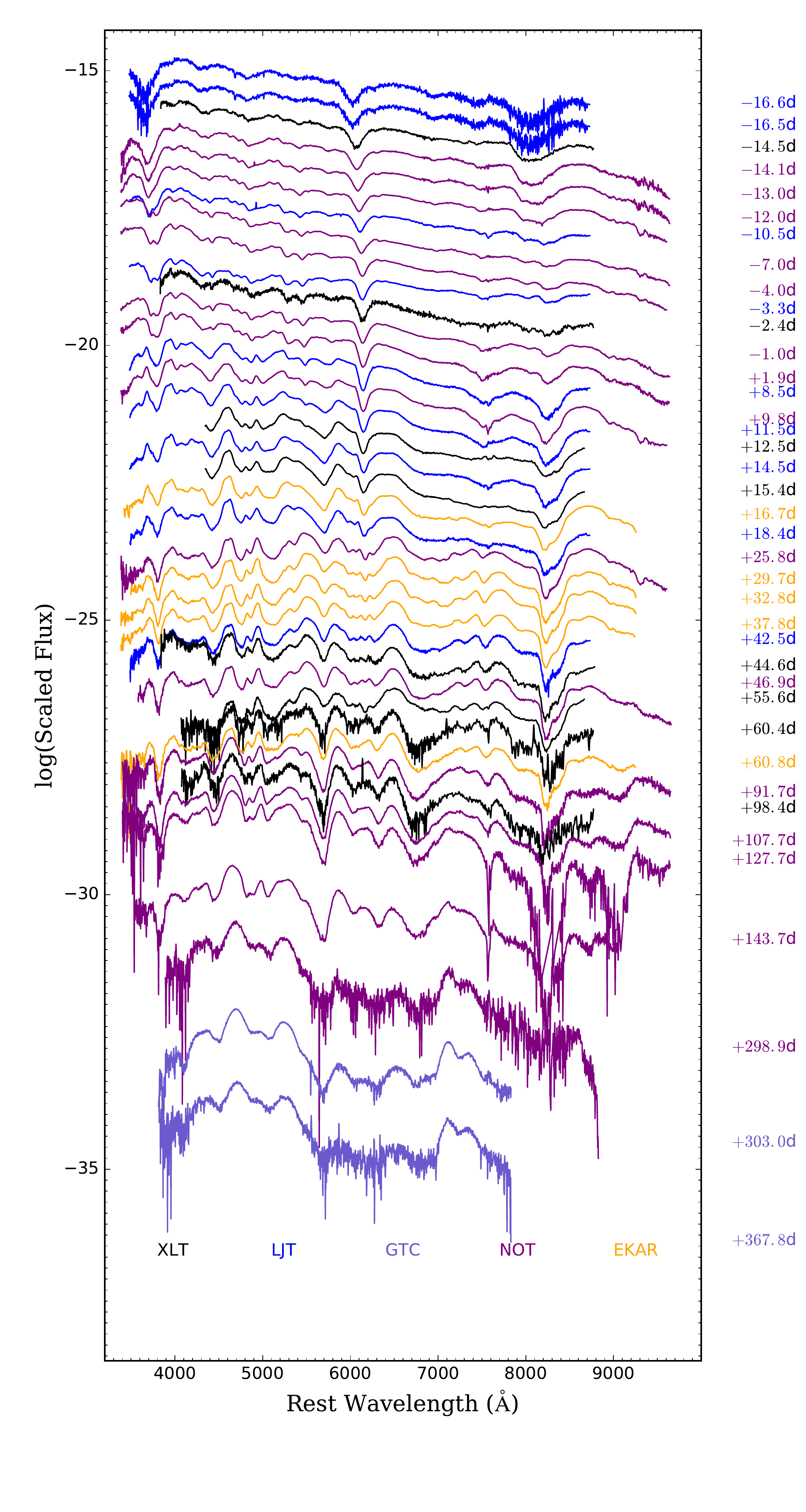}
    \caption{Optical spectral evolution of SN\,2019np from $-16.6$ days to $+367.8$ days relative to the B-band maximum light. The spectra have been corrected for host-galaxy redshift (z = 0.00452) and reddening. The text on the right side of each spectrum denotes the phase in days since the $B$-band maximum light. Different colors of the spectra represent that they were taken with different spectroscopic instruments (i.e., XLT, LJT, GTC, NOT or EKAR), which are indicated at the bottom of the plot.}
    \label{spectraall}
\end{figure*}

Figure~\ref{spectraall} shows the spectral evolution of SN\,2019np in the optical wavelength range.
The spectral evolution follows that of normal SNe\,Ia. For example, within the first weeks after the explosion, the spectra of SN\,2019np are characterized by prominent absorption lines of intermediate-mass elements (IMEs) and ionized IGEs. By the time SN\,2019np stepped into the early nebular phase, absorption features of IGEs start to dominate the spectra.

\begin{figure*}
    \centering
	\includegraphics[width =2.3\columnwidth]{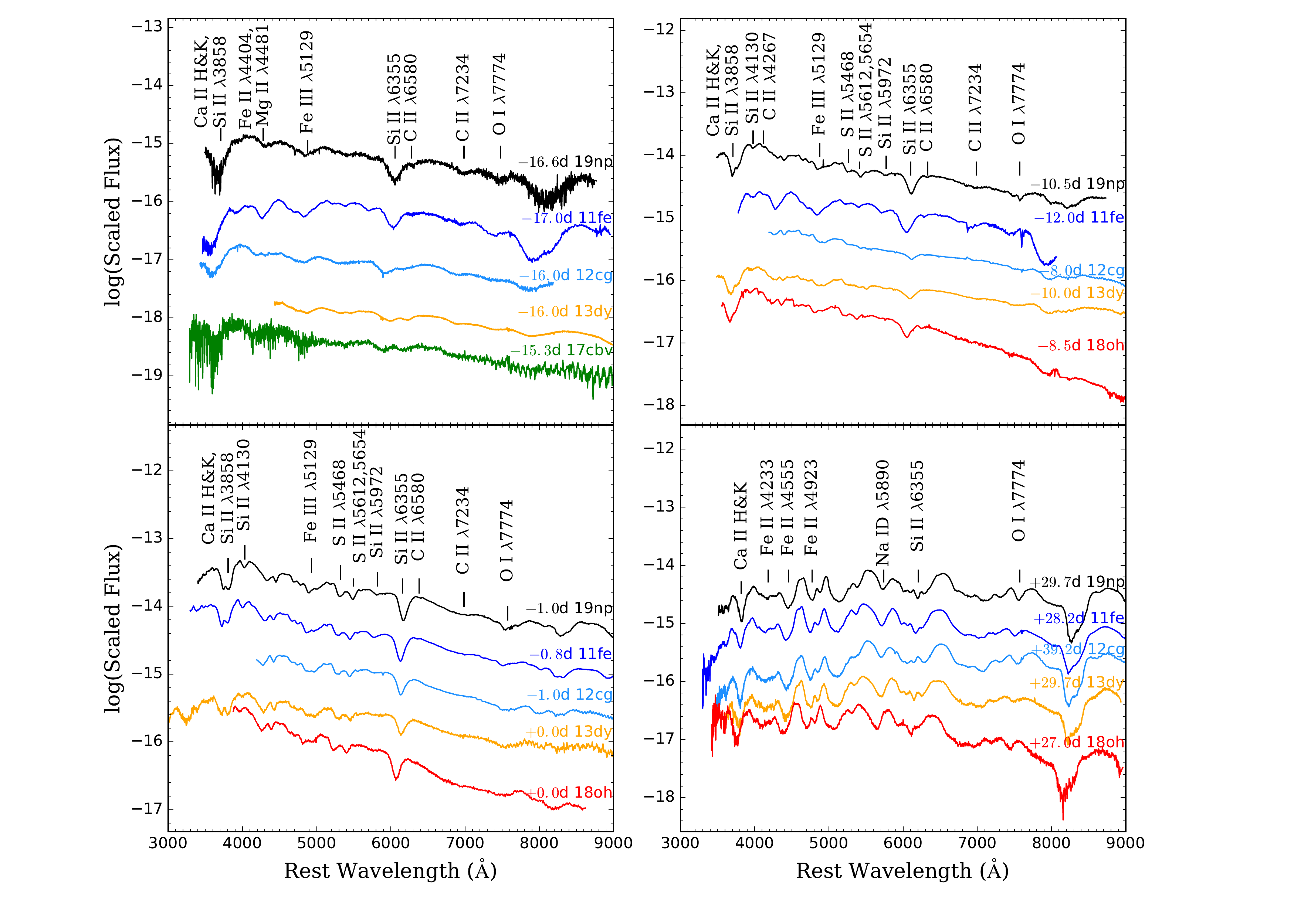}
    \caption{
    Spectral comparison between SN\,2019np and some well-studied SNe Ia at several selected epochs (i.e., t$\sim-16.0$ days, $-10.0$ days, $0$ days, and $+30.0$ days). The comparison objects include SNe 2011fe \citep{2014MNRAS.439.1959M,2016ApJ...820...67Z}, 2012cg \citep{2016ApJ...820...92M}, 2013dy \citep{2013ApJ...778L..15Z,2015MNRAS.452.4307P,2016AJ....151..125Z}, 2017cbv \citep{2017ApJ...845L..11H} and 2018oh \citep{2019ApJ...870...12L}. All spectra have been corrected for reddening and the host-galaxy redshifts. For better display, all spectra were shifted arbitrarily along the vertical direction.}
\label{spectra compare}
\end{figure*}

Figure~\ref{spectra compare} shows the spectral comparison between SN\,2019np and some well-observed normal SNe\,Ia with similar decline rates at different epochs. The earliest spectrum obtained at t$\sim -16.6$ d is dominated by prominent absorption lines of IMEs and IGEs. Note that weak absorption features of \ion{C}{ii} $\lambda$6580 and \ion{C}{ii} $\lambda$7234 can clearly be identified in its earliest spectra. The absorption feature near 4300 {\rm \AA} could be due to the \ion{Mg}{ii} $\lambda$4481 line blended with the \ion{Fe}{ii} $\lambda$4404 line, while the blended lines of \ion{Fe}{iii} $\lambda$5129, \ion{Fe}{ii} $\lambda\lambda\lambda$4924,5018,5169 and \ion{Si}{ii} $\lambda$5051 are responsible for the broad absorption near 4800 {\rm \AA}. The \ion{Ca}{ii} H\&K and \ion{Ca}{ii} NIR triplet contribute to the prominent absorption features near 3700 {\rm \AA} and 8000 {\rm \AA}, respectively. In the spectrum at t$\sim -10.5$ d spectrum, the ``W''-shaped \ion{S}{ii} absorption features near 5400 {\rm \AA} and the \ion{Si}{ii} $\lambda$5972 feature near 5800 {\rm \AA} start to show up in the spectra and gain their strengths. By this phase, the \ion{C}{ii} $\lambda$6580 absorption is still visible in SN 2019np, SN 2011fe, and SN 2018oh, but it disappears in SN 2012cg and SN 2013dy. Detached high velocity feature (HVF) can be clearly seen in \ion{Ca}{ii} NIR triplet of SN 2019np, which is similar to the comparison SNe Ia. The spectra around the maximum light are characterized by the prominent absorption features of \ion{Ca}{ii} H\&K, \ion{S}{ii}, \ion{Si}{ii} $\lambda$6355 and \ion{Ca}{ii} NIR triplet. Note that the \ion{C}{ii} $\lambda$6580 absorption feature still seems visible in SN 2019np, SN 2011fe and SN 2018oh. The HVF of the \ion{Ca}{ii} NIR triplet becomes weak while the photospheric component gain its strength.  At t$\sim1$ month, the characteristic \ion{Fe}{ii} features within the wavelength range of $4600-5300$\AA\ start to develop and dominate in the spectra. At this phase, the overall spectral features of SN\,2019np and the comparison SNe\,Ia become quite similar. Spectroscopically, SN 2019np is found to be more similar to SN 2011fe.

\begin{figure*}
	\includegraphics[scale=0.6]{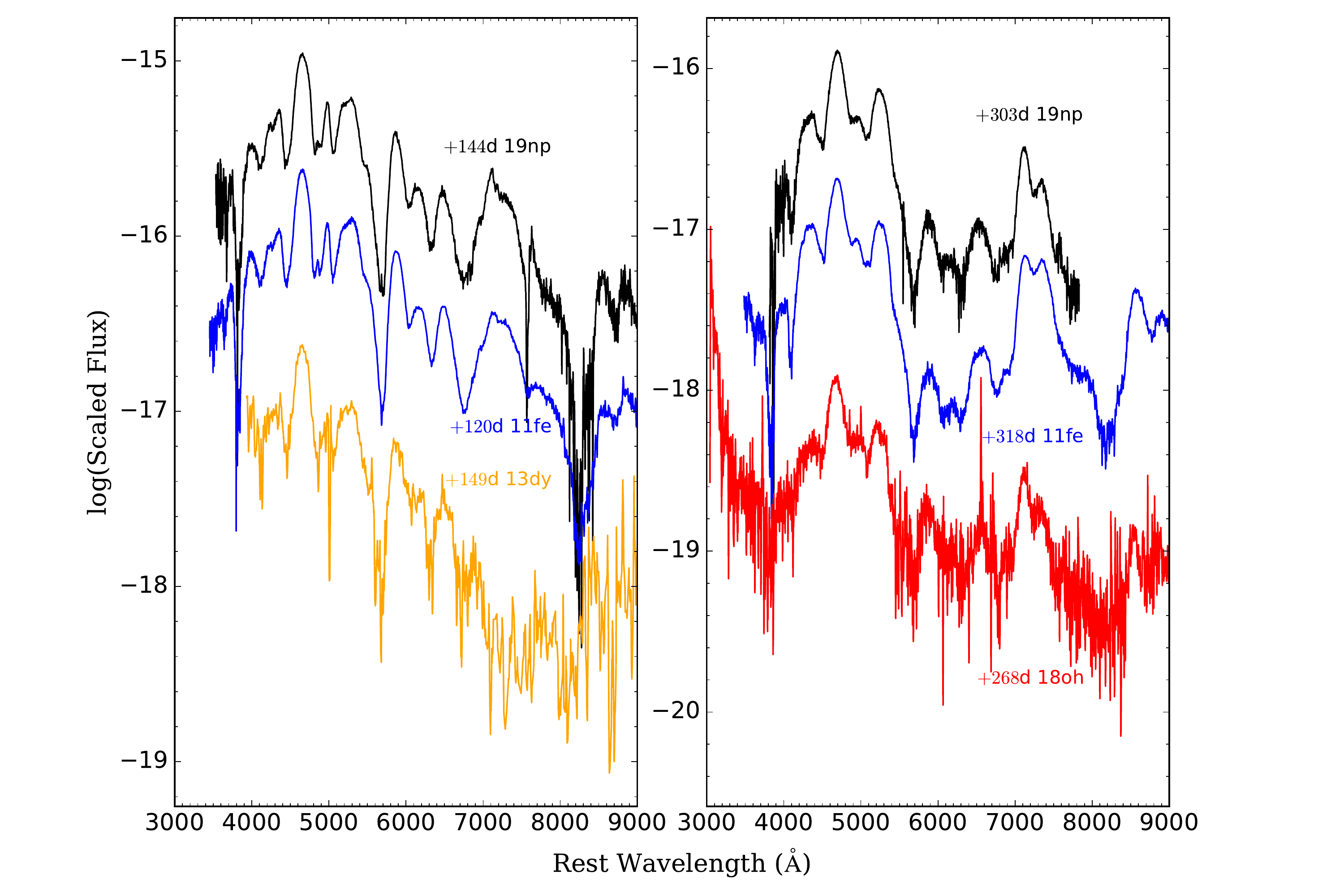}
    \caption{Nebular-phase spectra of SN\,2019np obtained at t$\sim+144$ d (left panel) and $+303$ d (right panel) compared with the spectra of a few well-observed SNe Ia at similar phases, including SNe\,2011fe \citep{ 2012MNRAS.425.1789S, 2012PASP..124..668Y}, 2013dy \citep{2013ApJ...778L..15Z,2016AJ....151..125Z} and 2018oh \citep{2019ApJ...872L..22T}. All spectra were corrected for host-galaxy redshifts. The yellow and red spectra have been smoothed by a Savitsky–Golay filter.}
     \label{latespectracompare}
\end{figure*}

Figure~\ref{latespectracompare} shows the spectra taken at t$\sim$144 d and t$\sim$303 d after the maximum light, when forbidden lines of ionized IGEs such as Fe and Co dominate in the spectra. One can see that the late-time behaviour of SN\,2019np also exhibits considerable similarities to that of SNe\,2011fe, 2013dy, and 2018oh. Combining the spectra at t $\sim$144 d and $\sim$303 d, we notice that the [Fe III] feature at $\sim4700$\AA\, [Co II] at $\sim6000$\AA\, and  [Fe II]/[Ni II] at $\sim7200$\AA\ tend to become relatively weak with time. From the t$\sim$303 d spectrum, we measured the velocity shift of forbidden emission lines of [\ion{Fe}{ii}]$7155$\AA\ and [\ion{Ni}{ii}]$7378$\AA\ as $1550 \pm 140$ km s$^{-1}$, which is consistent with that of the comparison normal SNe\,Ia (i.e. SN 2011fe and SN 2018oh).

\subsection{Ejecta Velocity}

\begin{figure}
\includegraphics[width=\columnwidth]{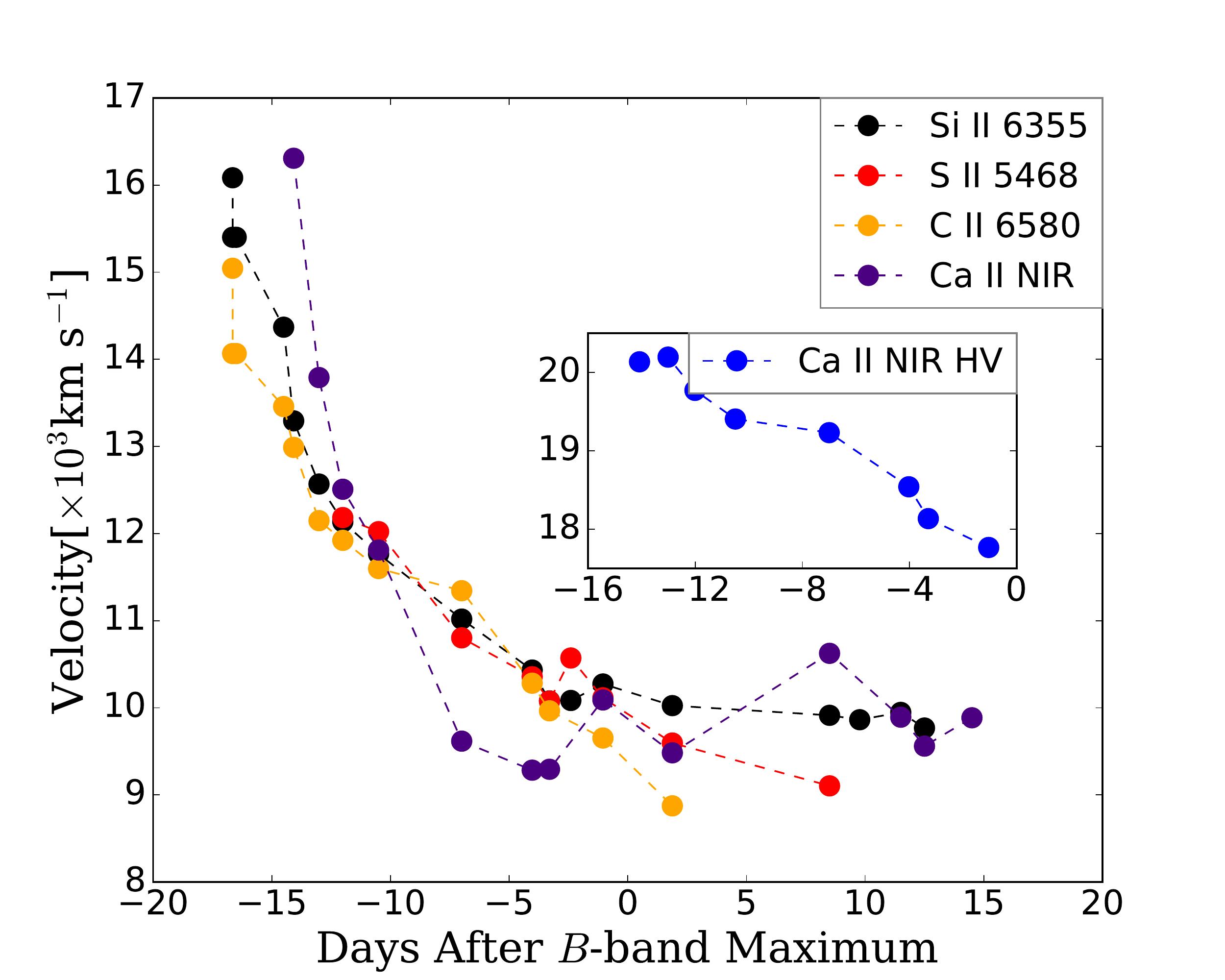}
 \caption{Evolution of the expansion velocity of SN\,2019np as measured from the absorption minimum of \ion{Si}{ii} $\lambda$6355, \ion{S}{ii} $\lambda$5468, \ion{C}{ii} $\lambda$6580 and \ion{Ca}{ii} NIR triplet.}
\label{velocity}
\end{figure}

We measure the ejecta velocities of SN\,2019np using the absorption minima of a series of spectral features including \ion{Si}{ii} $\lambda$6355, \ion{S}{ii} $\lambda$5468, \ion{C}{ii} $\lambda$6580, \ion{Ca}{ii} NIR triplet, and the result is shown in Figure~\ref{velocity}. All velocities have been corrected to the restframe of the host galaxy based on its redshift. 
The photospheric velocity of \ion{Si}{ii} $\lambda$6355 at t$\sim$-16.6 d is $\sim$16,100 km s$^{-1}$, which is higher compared to that measured from the \ion{C}{ii} $\lambda$6580 velocity of $\sim$15,000 km s$^{-1}$.

The velocity of \ion{Si}{ii} $\lambda$6355 is measured as 10,200 km s$^{-1}$ at the time of $B$-band maximum, which is comparable to the typical value (i.e., $\sim$10,500 km s$^{-1}$) of normal SNe\,Ia and can be clearly put into the normal velocity group according to the classification scheme proposed by \cite{2009ApJ...699L.139W}. The velocity of \ion{Si}{ii} $\lambda$6355 exhibits a similar evolution as that measured from the \ion{C}{ii} $\lambda$6580 and \ion{S}{ii} $\lambda$5468.

\begin{figure}
	\includegraphics[width=\columnwidth]{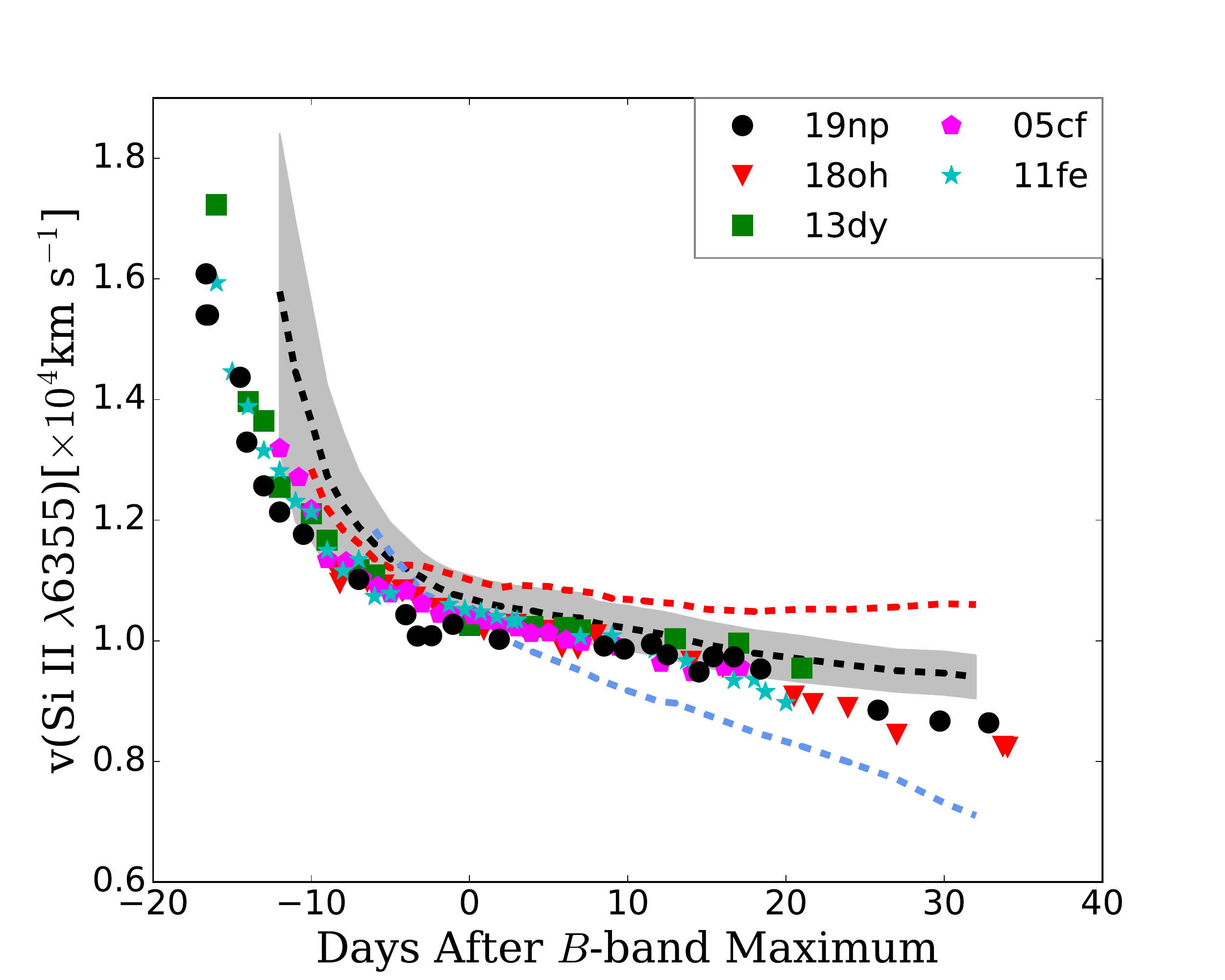}
    \caption{Velocity evolution of SN\,2019np as measured from the absorption minimum of \ion{Si}{ii} $\lambda$6355, compared to that of SNe\,2018oh, 2013dy, 2005cf and 2011fe. Overplotted curves present the velocity evolution obtained for SN\,1991T-like (red dashed), SN\,1991bg-like (blue dotted), and normal (solid black) subclasses of SNe\,Ia \citep{2009ApJ...699L.139W}. The gray-shaded region indicates the 1-$\sigma$ standard deviation of the mean velocity of normal SNe\,Ia. }
\label{sivelocity}
\end{figure}

Basic photometric and spectroscopic parameters of SN\,2019np are listed in Table \ref{tabpars}. The velocity evolution of SN\,2019np as derived from the \ion{Si}{ii} $\lambda$6355 line is presented in Figure~\ref{sivelocity}. The velocity evolution of SNe\,2018oh, 2013dy, 2005cf \citep{2009ApJ...697..380W} and 2011fe are also shown for comparison. 
The velocity gradient of \ion{Si}{ii} $\lambda$6355 measured 10 days past maximum is $\dot{v}_{\rm{Si}}$=\VSiIIDot, indicating that SN\,2019np belongs to the low-velocity gradient (LVG) subclass according to the classification scheme proposed by \cite{2005ApJ...623.1011B}.

\begin{table}
\centering
\caption{Parameters of SN\,2019np}
\label{tabpars}
\begin{tabular*}{3.0in}{@{\extracolsep{1in}}ll}
\hline\hline
\multicolumn{1}{c}{Parameter} & \multicolumn{1}{c}{Value} \\\hline
\multicolumn{2}{c}{Photometric}                            \\
$B_{\rm max}$                 & \mbobs\,mag  \\
$B_{\rm max}-V_{\rm max}$     & $0.005 \pm 0.007$\,mag \\
$M_{\rm max}(B)$              & \mbmag\,mag \\
$E(B-V)_{\rm host}$           & \ebvalue\,mag   \\
$\Delta m_{\rm15}(B)$            & \dmvalue\,mag \\
$s_{BV}$ 	                   & \sbvalue\     \\
$t_{\rm max}(B)$ 	         & \tbmax\,d \\
$t_0$	                         &	\tzero\,d \\
$\tau_{\rm rise}$	         & \trise\,d \\
$L_{\rm bol}^{\rm max}$	    & \Lmax\ \\
$M_{^{56}\rm Ni}$	          & \MniValue\ \\
$\mu$                              & \distmd\,mag \\
\multicolumn{2}{c}{Spectroscopic}                          \\
$v \rm _{0}$(\SiII)	          & \VSiII\ \\
$\dot{v}$(\SiII)		     & \VSiIIDot\ \\
$R$(\SiII)			          & \RSiII\ \\
\hline
\end{tabular*}
\end{table}

\subsection{NIR Spectra}
\begin{figure*}
\centering
	\includegraphics[width=2\columnwidth]{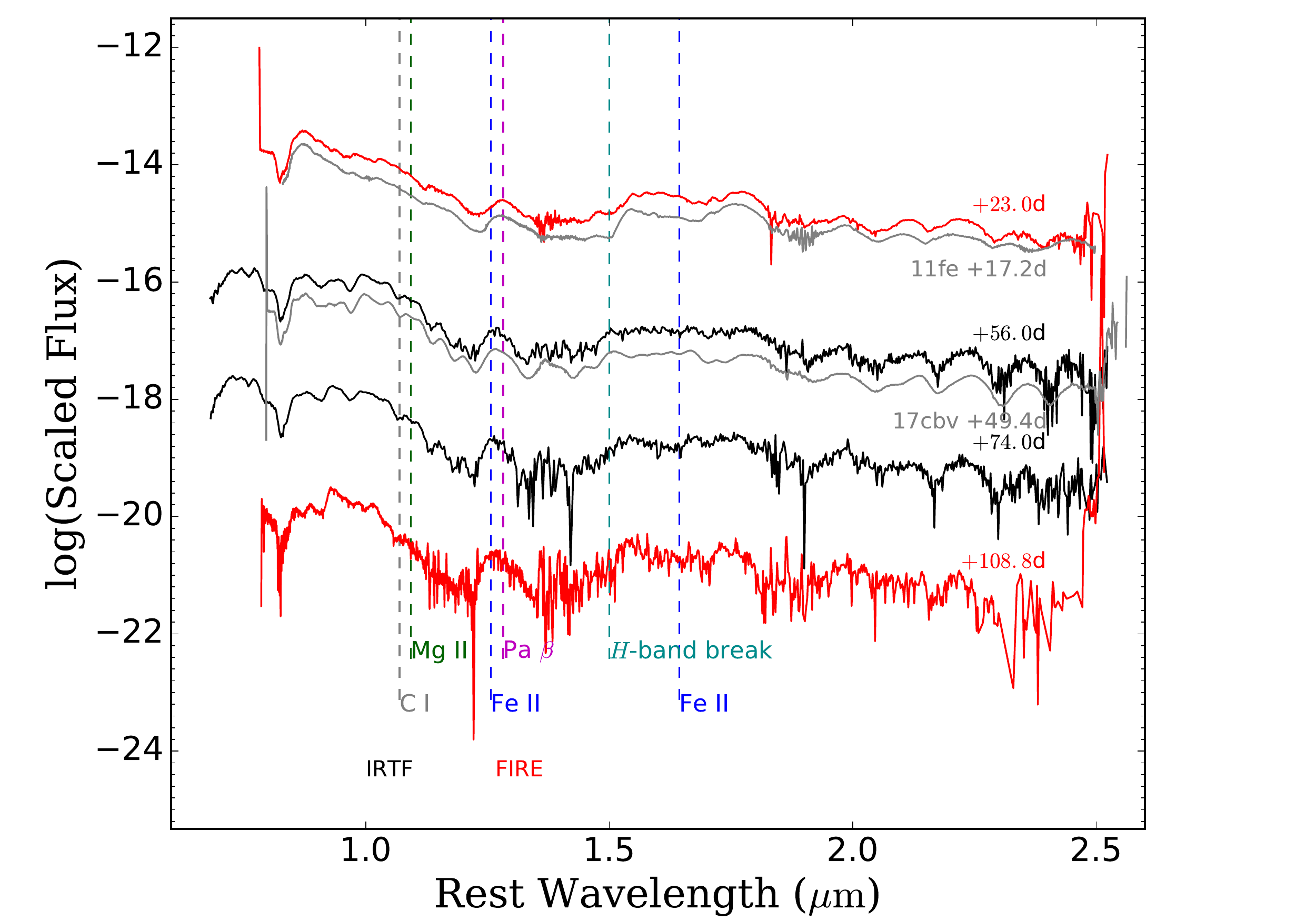}
    \caption{NIR spectra of SN\,2019np from FIRE and IRTF, along with the comparable-phase spectra of SN\,2011fe \citep{2013ApJ...766...72H} and SN\,2017cbv \citep{2020ApJ...904...14W}.}
\label{NIR spec}
\end{figure*}
Figure~\ref{NIR spec} shows the NIR spectra of SN\,2019np taken with IRTF and FIRE. We also compared the t$=+23.0$ d and $+56.0$ d spectra with SN\,2011fe and SN\,2017cbv. 

The nondetection of the Pa$\beta$ feature indicates that the mass of hydrogen contained in companion star is less than 0.1 $M_{\sun}$ \citep{2020ApJ...904...14W}. The lack of hydrogen features in the nebular spectra is also in accord with a low hydrogen mass limit \citep{2018ApJ...863...24S}.

Moreover, the $H$-band break is observed around 1.5$\mu$m in the NIR spectra of SN\,2019np, which is related to the dramatic shift in the amount of line blanketing from IGEs \citep{1998ApJ...496..908W}. Similar to other SNe\,Ia \citep{2018ApJ...863...24S}, a decrease in strength of the $H$-band break can be identified in SN\,2019np during the phases from $t=+23.0$ to $+108.8$ d. Such a time evolution of the $H$-band break is related to the quantity of intermediate-mass elements, depending on different explosion scenarios, which can be further used to determine mass of $^{56}$Ni  \citep{2019PASP..131a4002H,2019ApJ...875L..14A}. 

\section{Discussion}\label{sec:dis}

\subsection{Quasi-bolometric Light Curve}\label{arnettmodel}
\begin{figure}
	\includegraphics[width=\columnwidth]{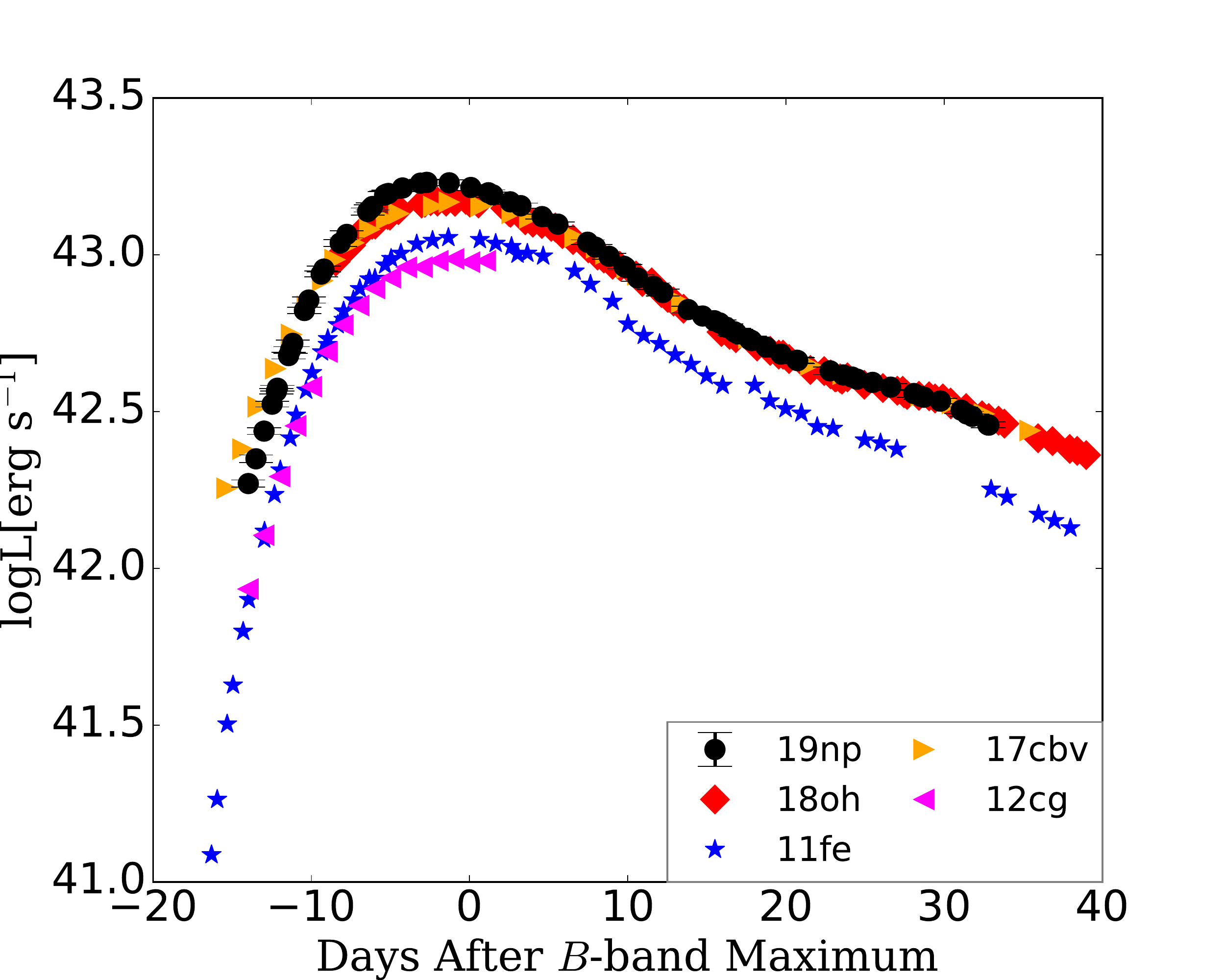}
    \caption{Quasi-bolometric light curve of SN\,2019np compared to that of SNe 2018oh, 2011fe, 2012cg, and 2017cbv.}
     \label{bolo}
\end{figure}

\begin{figure}
	\includegraphics[width=\columnwidth]{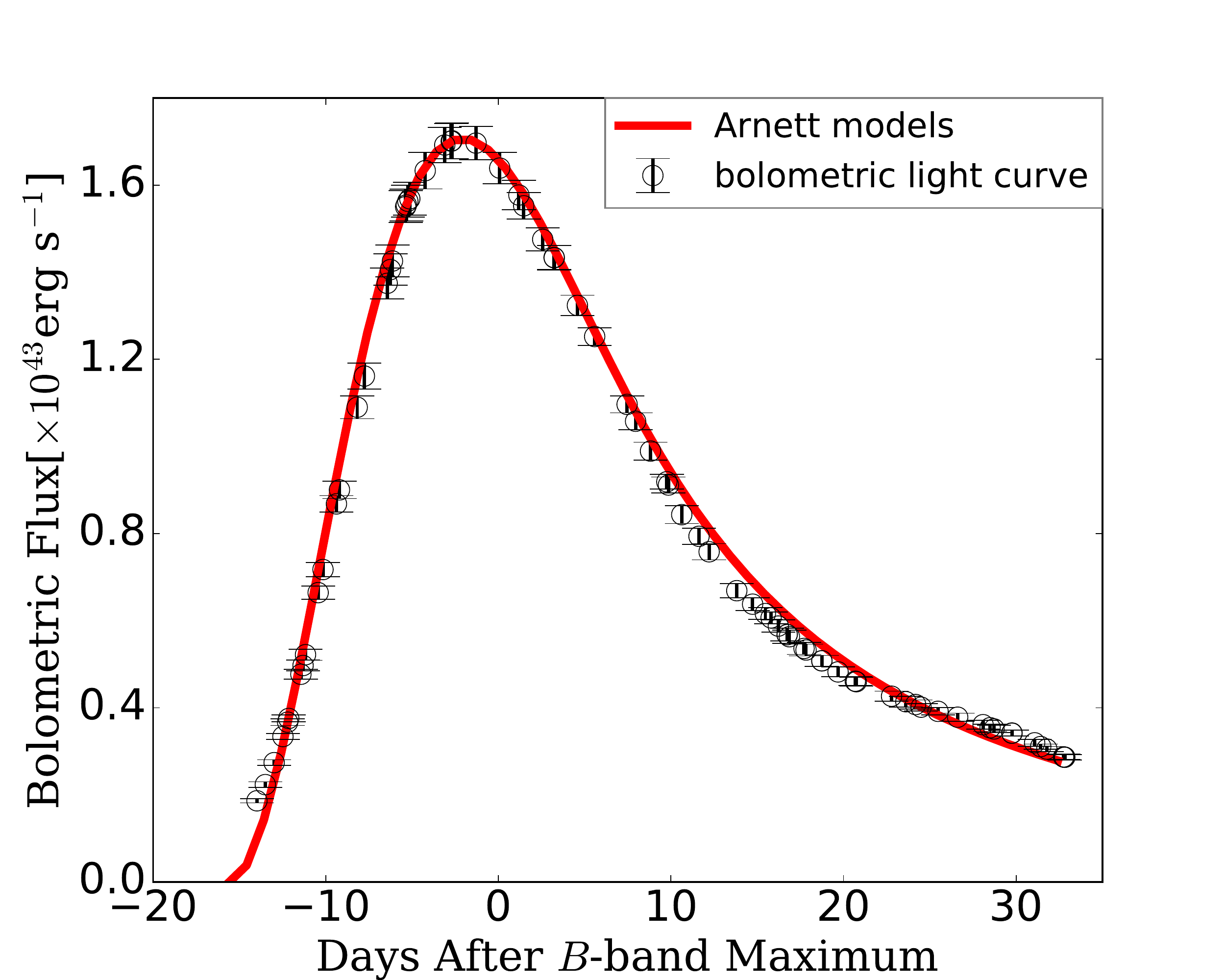}
    \caption{The quasi-bolometric light curve of SN\,2019np (open black circles) compared to the best-fit radiation diffusion model (red line; \citealt{1982ApJ...253..785A}).}
    \label{arnett}
\end{figure}

The distance of the host galaxy NGC 3254 is 32.80 $\pm$ 6.8 Mpc according to Tully-Fisher relation \citep{2016AJ....152...50T}. Assuming $R_V=3.1$, the absolute magnitude of $B$-band peak is estimated as $M_{\rm{max}}(B)=$\mbmag\,mag after correcting for the Galactic and host-galaxy extinctions. We derived the quasi-bolometric luminosity of SN\,2019np based on Swift UV and optical photometry. Assuming that the NIR-band emission of SN\,2019np has a similar contribution as in SN\,2011fe \citep{2016ApJ...820...67Z}, the quasi-bolometric light curve of SN\,2019np is shown in Figure~\ref{bolo}. The maximum-light luminosity is $L_{\rm peak} =$\Lmax, which is slightly larger than SN\,2011fe ($L_{\rm peak}^{\rm 11fe} = 1.13\times10^{43}$ erg s$^{-1}$, \citealt{2016ApJ...820...67Z}).
We use the {\tt Minim} Code \citep{2013ApJ...773...76C}, based on the radiation diffusion model from Arnett \citep{1982ApJ...253..785A,2012ApJ...746..121C,2019ApJ...870...12L, 2021ApJ...919...49Z}, to estimate the nickel mass and other parameters. The relevant result is shown in Figure~\ref{arnett}. The fitting parameters include first-light time $t_0=$\tzero, the mass of the radioactive nickel $M_{ \rm Ni} =$\MniValue, the light-curve timescale $t_{\rm lc}=$\tlcv\,days, and the gamma-ray leaking timescale $t_{\gamma}=$\tgama\,days \citep{2012ApJ...746..121C}. Then the inferred ejecta mass ($M_\mathrm{ej}$) and expansion velocity ($v_\textup{exp}$) can be calculated from $t_{\rm lc}$ and $t_{\gamma}$:
\begin{equation}
\centering
t_{\textup{lc}}^{2}=\frac{2\kappa M_{\textup{ej}}}{\beta c v_{\textup{exp}}}~\textup{and}~ t_{\gamma}^{2}=\frac{3\kappa_{\gamma}M_{\textup{ej}}}{4\pi v_\textup{exp}^{2} }   
\end{equation}
\citep{1982ApJ...253..785A,1997ApJ...491..375C,2008MNRAS.383.1485V,2012ApJ...746..121C,2015MNRAS.450.1295W,2019ApJ...870...12L}, where $\kappa$ is the effective opacity in optical, $\kappa_{\gamma}$ is the opacity for $\gamma$ rays (assuming that positrons released during cobalt decay are completely captured) and $\beta \sim 13.8$ is the light curve parameter related to the density profile of the ejecta \citep{1982ApJ...253..785A}. The $\kappa_{\gamma}$ can be estimated as $\kappa_{\gamma}\sim 0.03$ cm$^2$ g$^{-1}$ \citep{2015MNRAS.450.1295W}. So the $M_\mathrm{ej}$ and $v_\textup{exp}$ only depend on the value of the optical opacity $\kappa$. The $M_{ \rm ej}$ should be smaller than the Chandrasekhar mass (i.e., $M_{ \rm ej}<M_{ \rm ch}$) and the $v_\textup{exp}$ should be at least as large as the observed expansion velocity ($v_\textup{exp}> 11,000~{\rm km~s^{-1}}$). With these constraints, we derive $0.08\lesssim \kappa \lesssim0.09~{\rm cm^2~g^{-1}}$. Thus, we adopt $\kappa \sim 0.085~{\rm cm^2~g^{-1}}$ for SN 2019np (see also \cite{2019ApJ...870...12L} for a similar analysis of SN 2018oh). Then we get the ejecta mass $M_{ \rm ej}=1.34 \pm 0.12 M_{\sun}$ and the kinetic energy of the supernova as $E_\mathrm{sn}=0.3M_{ \rm ej}v_\textup{exp}^2=1.65^{+0.17}_{-0.15}\times 10^{51}$ erg.

\subsection{The Early-excess Flux}
 \begin{figure}
	\includegraphics[width=\columnwidth]{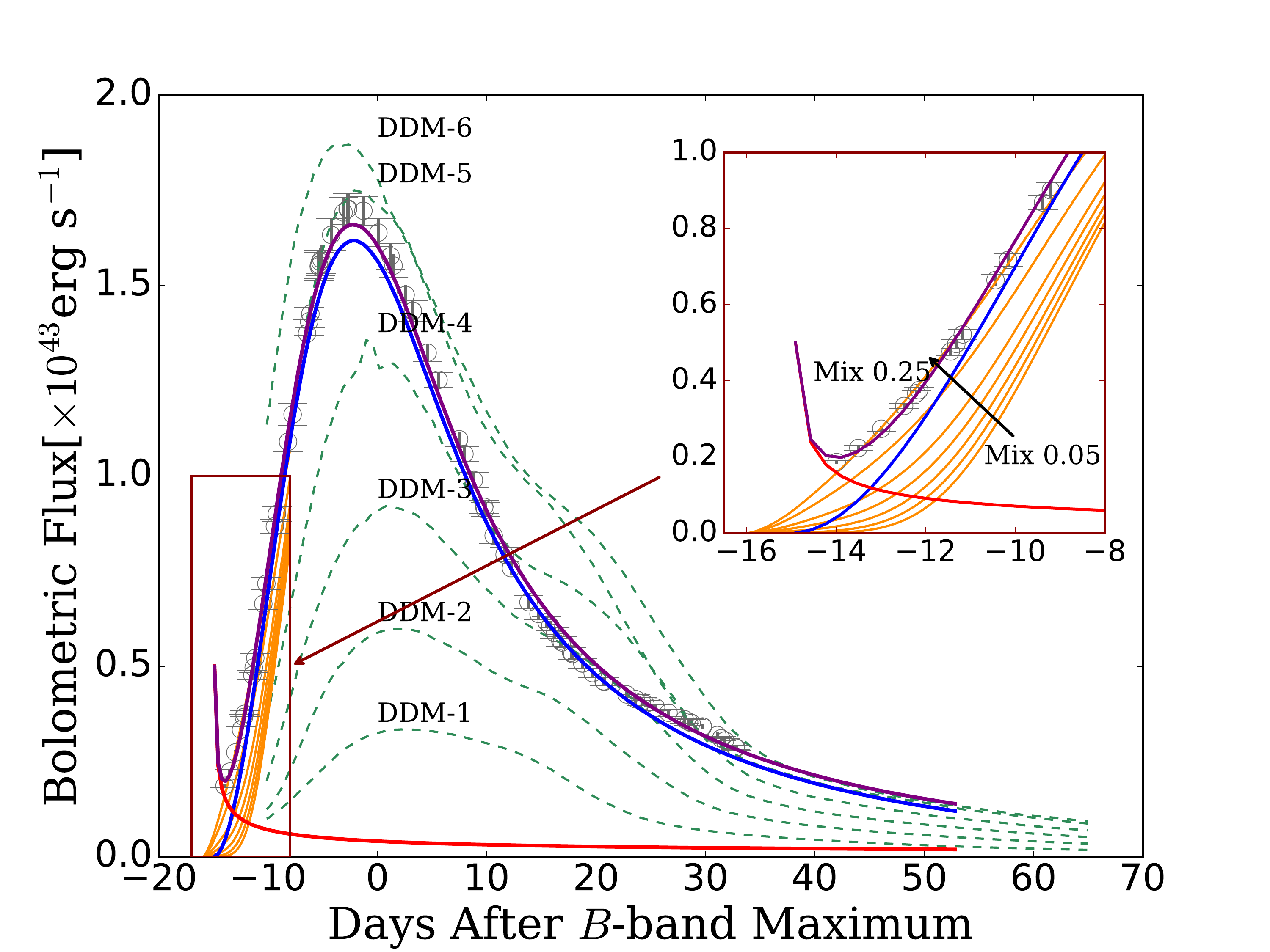}
    \caption{Bolometric light curve of SN\,2019np (grey open dots) compared with various models as indicated by different color-coded curves. The solid blue line represents the radioactive decay model, and the red solid line denotes the interaction model, while the purple solid line shows the sum of the above two. DDMs are shown in green dashed lines. Six models (DDM-1 to 6) are computed for an initial CO core of the WD and He shell masses of (0.810$M_{\sun}$, 0.126$M_{\sun}$), (0.920$M_{\sun}$, 0.084$M_{\sun}$), (1.025$M_{\sun}$, 0.055$M_{\sun}$), (1.125$M_{\sun}$, 0.039$M_{\sun}$), (1.280$M_{\sun}$, 0.013$M_{\sun}$), and (1.385$M_{\sun}$, 0.0035$M_{\sun}$), respectively. The inset portrays the data and the models within the first week as the region outlined by the red square. The orange lines denote the $^{56}$Ni mixing models; and the series of lines following the direction of the black arrow are generated using a boxcar with widths of 0.05, 0.075, 0.1, 0.125, 0.15, 0.2, and 0.25 $M_{\sun}$.}
    \label{bolo_model}
\end{figure}

\begin{figure}
	\includegraphics[width=\columnwidth]{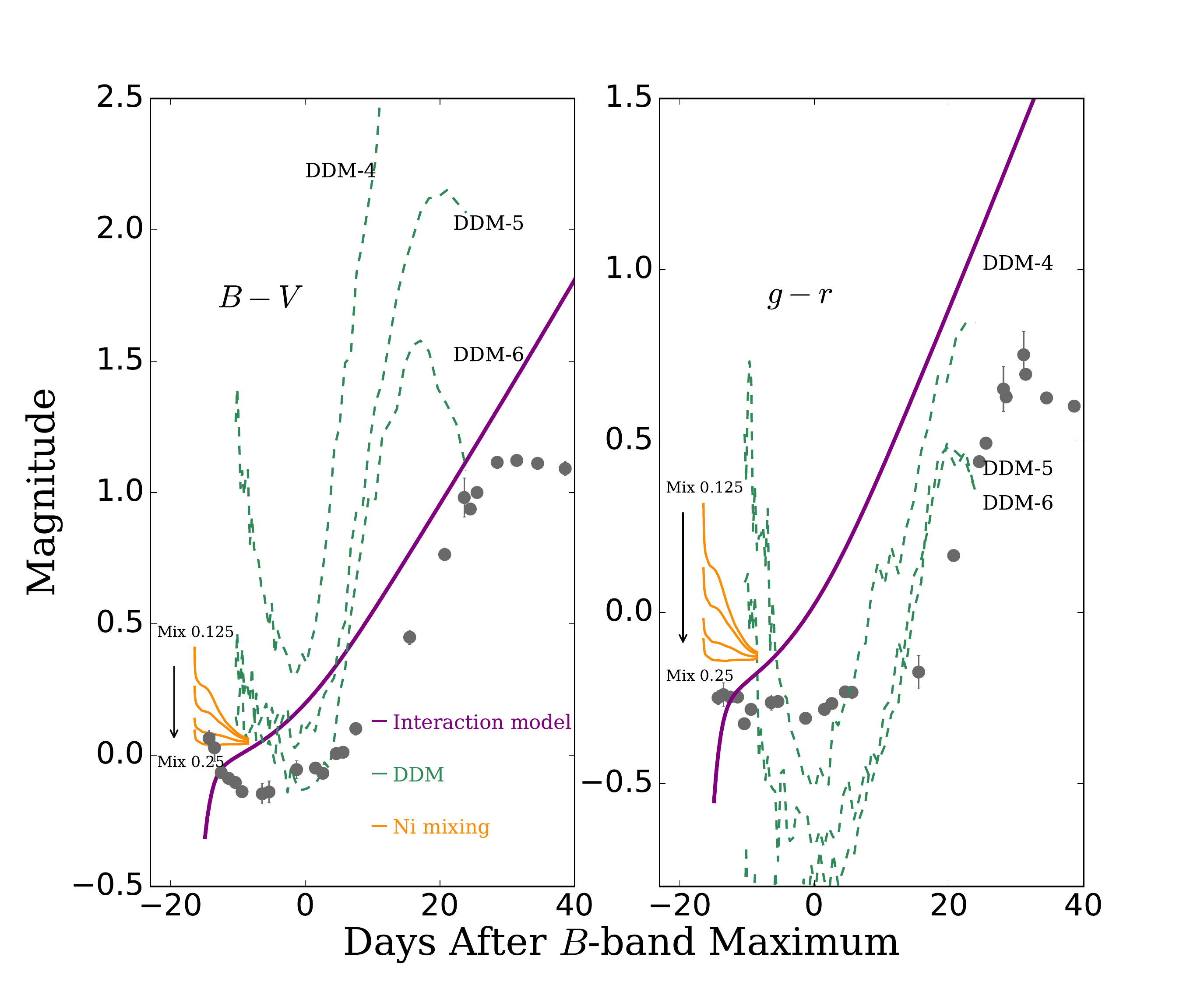}
    \caption{The $B-V$ bands (left panel) and $g-r$ bands (right panel) color of SN\,2019np (grey dots) compared with various explosion models as indicated by different color-coded curves. The models are shown in the same color with Figure~\ref{bolo_model}. }
    \label{color_model}
\end{figure}

In Figure~\ref{arnett} and Figure~\ref{uvotcompare}, one can see that excess flux exists in the early-time light curves. Such an excess emission is hard to be explained by diffusion of centrally-located $^{56}$Ni, but it could be related to the collision between the SN-ejecta and the companion star. Such a strong interaction would produce an optical/UV excess in the first week after the explosion. Thus, we use a hybrid model to fit the early-time light curves of SN\,2019np. Assuming that the density distribution of the ejecta follows a broken power law, i.e. $\rho\propto r^{-1}$ for the inner region and $\rho\propto r^{-10}$ for the outer one, the collision-powered SN luminosity ($L_\mathrm{col}$) and effective temperature ($T_\mathrm{col}$) can be expressed as follows \citep{2010ApJ...708.1025K}:
\begin{eqnarray}
L_\mathrm{col}&\propto&a M_\mathrm{ej}^{-5/8}E_\mathrm{sn}^{7/8}\kappa^{-3/4}t^{-1/2},\\
T_\mathrm{col}&\propto& a^{1/4} \kappa^{-35/144}t^{-37/72},
\end{eqnarray}
where $a$ is the distance between the mass center of the WD and its companion, and $t$ represents the dynamical time. We can then derive the time-dependent collision-powered flux ($F_{\nu,\mathrm{col}}$) under the assumption of blackbody radiation. For simplicity, the contribution from $^{56}$Ni-powered radiation is described by the fireball model ($F_{nu,\mathrm{Ni}}\propto t^2$; \citealp{2011Natur.480..344N}). So the total flux can be given by $F_{\nu}=F_{\nu,\mathrm{col}}+F_{\nu,\mathrm{Ni}}$.

In this circumstance, we still set $\kappa \sim 0.085~{\rm cm^2~g^{-1}}$ and adopt the above fitting results ($M_\mathrm{ej}$, $E_\mathrm{sn}$) from Arnett model. Then we fit the data observed earlier than MJD 58497 and derive the best-fit parameters, which suggests that the explosion date is MJD $58495.2\pm{0.1}$ and the separation distance between the WD and its companion star ($a$) is $5.74\pm {0.38}\times10^{11}$ cm. In most cases, the separation distance is comparable to the Roche-lobe radius of the companion star, for typical mass ratios, $a/R_{\rm companion}=2-3$ \citep{2010ApJ...708.1025K}. So the companion star could have a radius of $1.91-2.87\times 10^{11}$ cm. Thus, the corresponding companion star is likely to be a main-sequence star of $\sim1M_\odot$ (see Table 1 in \citealt{2010ApJ...708.1025K}). The best-fit result is presented in Figure~\ref{bolo_model}.

In addition, there are other possible explanations for the early bump in the light curves and one particular mechanism is the $^{56}$Ni mixing to the outermost layers \citep{2013ApJ...769...67P,2019ApJ...870L...1D}. In this scenario, the mass fraction of $^{56}$Ni near the surface exceeds that in the ``expanding fireball'' model, which can produce extra flux, and it can happen in both SD and DD progenitor channels. A specific explosion model for such a configuration is the double-detonation sub-Chandrasekhar explosion, where the surface $^{56}$Ni can be produced through the detonation of surface helium \citep{2017MNRAS.472.2787N}.

For the double detonation model (DDM), the explosion is thought to start with a shell of helium-rich material accreted by a sub-Chandrasekhar WD from its companion. The shell donation drives a shock wave that propagates towards the core of the CO WD, which could cause a secondary explosion in the core. The core explosion completely disrupts the WD \citep{1994ApJ...423..371W}. 
The DDMs 1-6, with parameters spanning from $0.810-1.385M_{\sun}$ WDs and the helium shells from $0.0035-0.126M_{\sun}$
\citep{2010ApJ...719.1067K,2010A&A...514A..53F}, are available from the Heidelberg Supernova Model Archive\footnote{https://hesma.h-its.org/}. We compared the bolometric light curves of SN\,2019np with those predicted by the above models in Figure~\ref{bolo_model}. 
The explosion time given by Arnett model is $t_0=$\tzero d. Note, however, that the peak bolometric luminosity predicted by DDMs 4-6 is higher than the observations. In addition, the DDMs 4-6 predict a shoulder in the bolometric light curve at around $\sim25.0$ day, which is not seen in SN\,2019np.

Furthermore, it is possible that $^{56}$Ni can also be mixed to the outer layer in some other scenarios \citep{2013ApJ...769...67P,2016ApJ...826...96P}. We consider a more general model in which the amount of surface $^{56}$Ni is tuned to explore the influence on the rising light curves \citep{2016ApJ...826...96P}. In Figure~\ref{bolo_model}, we present some typical light curves given by theoretical models \citep{2016ApJ...826...96P} which adopted a fixed amount of $^{56}$Ni (i.e., 0.5 $M_{\sun}$) and vary the distribution with a boxcar and a width from 0.05 to 0.25 $M_{\sun}$. As mentioned in \cite{2018ApJ...864L..35S} and \cite{2019ApJ...870L...1D}, the early-time color is an important parameter to distinguish the above models. In Figure~\ref{color_model}, we show the color evolution of SN\,2019np and the expected colors for different models. SN\,2019np shows broader red-blue-red (in $B-V$ color) evolution than the DDMs. Moreover, the $g-r$ color evolution of SN\,2019np is quite flat and does not conform with that predicted by the DDM. For the interaction model, the early-time color was initially blue and then became red, which is inconsistent with the observed color evolution of SN\,2019np. Based on the discussions above, we conclude that the early flux excess in SN\,2019np is most consistent with the $^{56}$Ni mixing model.

\section{Conclusion}\label{sec:con}
We present extensive UV-optical photometry and optical-NIR spectroscopy of the Type Ia SN\,2019np. The following conclusions can be made: 

1) The light curves suggest that SN\,2019np resembles that of other well-sampled normal SNe\,Ia. A $B-$band peak magnitude of \mbobs\, mag was reached, corresponding to an absolute magnitude of \mbmag\,mag. A $B-$band luminosity decline rate is found to be $\Delta m_{\rm15}(B)$ =\dmvalue\,mag. Both values are consistent with the
photometric behavior of normal SNe\,Ia; 
 
2) We construct the quasi-bolometric light curve of SN\,2019np based on the UV-optical photometry and an adoption of a NIR flux correction. The estimated peak bolometric luminosity gives $L_{\rm peak} =$\Lmax, indicating a synthesized nickel mass of \MniValue; 

3) Photometry of SN\,2019np started within the first two days of the explosion reveals an early blue bump in its lightcurves. The early bolometric flux evolution is up to $\sim$\excess higher compared to that predicted by the diffusion of internal radiation through a homogeneous expanding ejecta \citep{1982ApJ...253..785A}, suggesting the presence of energy sources in addition to the radioactive decay of nickel; 

4) We also compare the early color evolution of SN\,2019np to various models, including i) interaction between the SN ejecta and the CSM/companion star, ii) double-detonation explosion of a sub-Chandrasekhar mass WD, and iii) nickel mixed into the surface layers of the SN ejecta. We suggest that model iii) provides the most satisfactory fit; 

5) NIR and nebular-phase spectra of SN\,2019np show no evidence of a significant amount of Hydrogen, which is incompatible with the picture of the ejecta interacting with a nondegenerate companion or H-rich CSM. 

Early observations of SNe\,Ia, especially those with conspicuous flux excess that deviates from radiation diffusion through a homogeneously expanding ejecta within the first few days (``early-excess SNe\,Ia''), play an important role in understanding the SNe\,Ia explosion mechanism and the nature of their progenitors \citep{2018ApJ...865..149J,2018ApJ...864L..35S}. 
To our knowledge, only a handful of SNe\,Ia have been reported to show such an early excess, e.g., SNe\,2012cg \citep{2016ApJ...820...92M}, 2013dy \citep{2013ApJ...778L..15Z,2016AJ....151..125Z}, 2017cbv \citep{2017ApJ...845L..11H,2020ApJ...904...14W}, 2018oh \citep{2019ApJ...870L...1D}, iPTF13dge \citep{2016A&A...592A..40F}. The ejecta of the SN\,Ia expand rapidly and wipe out almost all traces of the pre-explosion configuration within days after the explosion of the progenitor WD. Therefore, photometry and spectroscopy starting within days after the explosion provides a unique view of the kinematics and mean chemical composition of the outermost layers of the SN ejecta and their circumstellar environments. High-cadence monitoring is also essential to probe these properties from the outer to inner parts of the ejecta as the photosphere continuously recedes into the exploding WD. Facilitated by the alert streams of modern wide-field, high-cadence transient searches, such as the Zwicky Transient Facility \citep{2019PASP..131a8002B,2019PASP..131f8003B}, ATLAS \citep{2011PASP..123...58T}, ASAS-SN \citep{2014AAS...22323603S}, DLT40 \citep{2018ApJ...853...62T} and TMTS \citep{2022MNRAS.509.2362L}, rapid photometric and spectroscopic follow-up observations will provide a more comprehensive characterization of the early behaviors of SNe\,Ia. An extensive sample of early SNe\,Ia will enable stringent tests among various models that may account for the flux excess in their early luminosity evolution. It will also provide chances to test the aspect-angle dependency and uniformities of SNe\,Ia that belong to different subtypes.

\section*{Acknowledgements}
We thank the anonymous referee for his/her constructive comments which help improve the manuscript. We are grateful to the staffs of the various telescopes and observatories with which data were obtained (Tsinghua-NAOC Telescope, Lijiang Telescope, Xinglong 2.16~m Telescope, Yunnan Astronomical Observatory and the 10.4~m Gran Telescopio CANARIAS). Financial support for this work has been provided by the National Science Foundation of China (NSFC grants 12033003 and 11633002), the Scholar Program of Beijing Academy of Science and Technology (DZ:BS202002), and the Tencent XPLORER Prize. This work was partially supported by the Open Project Program of the Key Laboratory of Optical Astronomy, National Astronomical Observatories, Chinese Academy of Sciences.

J.Z. is supported by the National Natural Science Foundation of China (NSFC, grants 11773067, 12173082, 11403096), by the Youth Innovation Promotion Association of the CAS (grant 2018081), and by the Ten Thousand Talents Program of Yunnan for Top-notch Young Talents. A.R. acknowledges support from ANID BECAS/DOCTORADO NACIONAL 21202412. Based in part on observations collected at Copernico and Schmidt telescopes (Asiago, Italy) of the INAF – Osservatorio Astronomico di Padova. M.D.S. is supported by grants from the VILLUM FONDEN (grant number 28021) and  the Independent Research Fund Denmark (IRFD; 8021-00170B). NUTS2’s use of  the Nordic Optical Telescope (NOT) is funded partially by the Instrument Center for Danish Astrophysics (IDA). SY acknowledge support from the G.R.E.A.T research environment, funded by {\em Vetenskapsr\aa det}, the Swedish Research Council, project number 2016-06012. Y.-Z. Cai  is funded by China Postdoctoral Science Foundation (grant no. 2021M691821). This work has been supported by MINECO grant ESP2017-82674-R , by EU FEDER funds and by grants 2014SGR1458 and CERCA Programe of the Generalitat de Catalunya (JI). This work made use of the Heidelberg Supernova Model Archive (HESMA), https://hesma.h-its.org. M.S. acknowledges the Infrared Telescope Facility, which is operated by the University of Hawaii under contract 80HQTR19D0030 with the National Aeronautics and Space Administration. The research of Y.Y. is supported through the Bengier-Winslow-Robertson Fellowship. M. Stritzinger is supported by grants from the VILLUM FONDEN (grant number 28021) and  the Independent Research Fund Denmark (IRFD; 8021-00170B). Lingzhi Wang is sponsored (in part) by the Chinese Academy of Sciences (CAS), through a grant to the CAS South America Center for Astronomy (CASSACA) in Santiago, Chile. CYW is supported by the National Natural Science Foundation of China (NSFC grants 12003013). BW is supported by the National Key R\&D Program of China (No. 2021YFA1600404), the Western Light Project of CAS (No. XBZG-ZDSYS-202117), the science research grants from the China Manned Space Project (No CMS-CSST-2021-A13).

\section*{Data availability}
The data underlying this article are available in the article. Our photometric data has been attached in Table~\ref{standstar},~ \ref{swift} and \ref{tab:optical} in the appendix. The log tables of optical and NIR spectral data are also shown in Table~\ref{log} and Table~\ref{lognir}.

\bibliographystyle{mnras}
\bibliography{2019np.bib} 

\section*{Affiliations}
\noindent
\textit{
$^{12}$Millennium Institute of Astrophysics, Nuncio Monsenor S\'{o}tero Sanz 100, Providencia, Santiago 8320000, Chile\\
$^{13}$George P. and Cynthia Woods Mitchell Institute for Fundamental Physics \& Astronomy, Texas A. \& M. University, 4242 TAMU, College Station, TX 77843, USA\\
$^{14}$Department of Physics, Florida State University, 77 Chieftan Way, Tallahassee, FL 32306, USA\\
$^{15}$Institut d'Estudis Espacials de Catalunya (IEEC), c/Gran Capit\'a 2-4, Edif. Nexus 201, 08034 Barcelona, Spain\\
$^{16}$Itagaki Astronomical Observatory, Yamagata, Yamagata 990-2492, Japan\\
$^{17}$The School of Physics and Astronomy, Tel Aviv University, Tel Aviv 69978, Israel\\
$^{18}$Key Laboratory of Optical Astronomy, National Astronomical Observatories, Chinese Academy of Sciences, Beijing 100101, China\\
$^{19}$School of Astronomy and Space Science, University of Chinese Academy of Sciences, Beijing 101408, China\\
$^{20}$Facultad de Ciencias Astron\'omicas y Geof\'isicas, Universidad Nacional de La Plata, Paseo del Bosque S/N, B1900FWA, La Plata, Argentina\\
$^{21}$Carnegie Observatories, Las Campanas Observatory, Colina El Pino, Casilla 601, Chile\\
$^{22}$Max-Planck-Institut f\"ur Astrophysik, Karl-Schwarzschild Str.~1, 85741 Garching, Germany\\
$^{23}$Technische Universit\"at M\"unchen, Physik Department, James-Franck Str. 1, 85741 Garching, Germany\\
$^{24}$Department of Physics and Astronomy, Aarhus University, Ny Munkegade 120, DK-8000 Aarhus C, Denmark\\
$^{25}$Chinese Academy of Sciences South America Center for Astronomy (CASSACA), National Astronomical Observatories, CAS, Beijing 100101,\\ People’s Republic of China\\
$^{26}$CAS Key Laboratory of Optical Astronomy, National Astronomical Observatories, Chinese Academy of Sciences, Beijing 100101, People’s Republic of China\\
$^{27}$Department of Astronomy, The Oskar Klein Center, Stockholm University, AlbaNova, 10691 Stockholm, Sweden\\
$^{28}$Department of Astronomy, Beĳing Normal University, Beĳing, 100875, People’s Republic of China\\}

\appendix
\section{PHOTOMETRIC AND SPECTROSCOPIC DATA}
\begin{table*}
  \centering
  \caption{Photometric Standards in the SN\,2019np Field for \textit{BVgri} bands}
    \begin{tabular}{cccccccccccccc}
\hline
Num. &$\alpha$(J2000) &
$\delta$(J2000) & \textit{B} (mag) & \textit{V} (mag) & \textit{g} (mag) & \textit{r} (mag) & \textit{i} (mag) \\	
\hline
    1     & $10^\mathrm{h}29^\mathrm{m}06^\mathrm{s}.86$ & $29\degr30\arcmin01\arcsec.81$ & 14.826(035) & 14.167(012) & 14.375(003) & 13.969(002) & 13.845(004) \\
    2     & $10^\mathrm{h}29^\mathrm{m}45^\mathrm{s}.12$ & $29\degr33\arcmin23\arcsec.79$ & 14.446(036) & 13.824(013) & 14.014(007) & 13.641(001) & 13.527(013) \\
    3     & $10^\mathrm{h}29^\mathrm{m}16^\mathrm{s}.58$ & $29\degr30\arcmin28\arcsec.17$ & 16.456(035) & 15.745(013) & 15.977(005) & 15.524(002) & 15.348(003) \\
    4     & $10^\mathrm{h}29^\mathrm{m}24^\mathrm{s}.40$ & $29\degr29\arcmin31\arcsec.28$ & 16.825(035) & 15.862(013) & 16.213(005) & 15.534(003) & 15.272(002) \\
    5     & $10^\mathrm{h}29^\mathrm{m}17^\mathrm{s}.43$ & $29\degr25\arcmin47\arcsec.52$ & 17.187(035) & 16.279(014) & 16.604(006) & 15.974(004) & 15.756(003) \\
    6     & $10^\mathrm{h}29^\mathrm{m}05^\mathrm{s}.83$ & $29\degr28\arcmin26\arcsec.37$ & 16.281(035) & 15.674(013) & 15.857(004) & 15.498(002) & 15.390(003) \\
    7     & $10^\mathrm{h}29^\mathrm{m}22^\mathrm{s}.53$ & $29\degr34\arcmin07\arcsec.31$ & 18.035(035) & 16.958(014) & 17.363(006) & 16.581(004) & 16.283(004) \\
    8     & $10^\mathrm{h}29^\mathrm{m}21^\mathrm{s}.14$ & $29\degr33\arcmin19\arcsec.29$ & 17.820(035) & 16.962(013) & 17.264(005) & 16.679(003) & 16.456(002) \\
    9     & $10^\mathrm{h}29^\mathrm{m}33^\mathrm{s}.57$ & $29\degr34\arcmin21\arcsec.00$ & 18.793(038) & 17.996(021) & 18.269(010) & 17.739(011) & 17.451(005) \\
    10    & $10^\mathrm{h}29^\mathrm{m}19^\mathrm{s}.68$ & $29\degr32\arcmin25\arcsec.88$ & 19.127(035) & 17.578(014) & 18.206(006) & 17.000(004) & 15.672(002) \\
    11    & $10^\mathrm{h}29^\mathrm{m}33^\mathrm{s}.65$ & $29\degr25\arcmin24\arcsec.47$ & 18.281(035) & 17.325(014) & 17.673(006) & 17.000(004) & 16.748(005) \\
\hline
\multicolumn{5}{l}{Note: Uncertainties, in units of 0.001 mag, are $1\sigma$.}\\
    \end{tabular}%
  \label{standstar}%
\end{table*}%

\begin{table*}
\centering
\caption{\textit{Swift} Photometry of SN\,2019np}
\begin{tabular}{ccccccccc}
\hline
MJD &\textit{uvw2} (mag) & \textit{uvm2} (mag) & \textit{uvw1} (mag) & \textit{UVOT u} (mag)& \textit{UVOT b} (mag)& \textit{UVOT v} (mag)\\
\hline
    58493.7  & 19.644(315) & ... & 18.868(281) & 17.091(107) & 16.979(084) & 16.868(123) \\
    58494.5  & 19.703(255) & ... & 18.576(170) & 16.859(085) & 16.586(067) & 16.415(086) \\
    58495.7  & 19.156(182) & ... & 17.832(115) & 16.295(073) & 15.959(060) & 15.742(069) \\
    58497.9  & 18.057(114) & ... & 16.625(078) & 14.974(055) & 15.014(048) & 15.012(066) \\
    58499.7  & 17.125(084) & 18.817(161) & 15.610(063) & 14.022(044) & 14.473(043) & 14.460(054) \\
    58501.9  & 16.496(072) & 17.880(104) & 14.860(057) & 13.300(041) & 13.968(041) & 14.003(048) \\
    58502.4  & 16.334(068) & 17.881(100) & ... & ... & ... & ... \\
    58503.2  & 16.216(069) & 17.552(093) & 14.579(055) & 13.041(040) & 13.781(041) & 13.805(046) \\
    58507.8  & 15.925(075) & 16.916(091) & 14.318(055) & 12.766(040) & 13.443(040) & 13.479(045) \\
    58509.9  & 15.983(071) & 17.058(086) & 14.464(056) & 12.896(040) & 13.374(040) & 13.385(044) \\
    58511.0  & 16.037(072) & 17.038(086) & 14.536(057) & 13.030(040) & 13.433(040) & 13.378(043) \\
    58522.8  & 17.244(104) & 18.048(138) & 16.061(079) & 14.500(050) & 14.193(043) & 13.783(047) \\
    58525.7  & 17.653(137) & 18.534(201) & 16.503(101) & 14.857(058) & 14.475(045) & 13.985(051) \\
    58528.2  & 17.824(124) & 18.651(179) & 16.629(090) & 15.168(062) & 14.800(047) & 14.141(051) \\
    58531.6  & 18.279(156) & 19.610(335) & 17.070(109) & 15.699(076) & 15.170(051) & 14.263(054) \\
    58537.4  & 18.889(166) & 19.647(263) & 17.718(115) & 16.207(073) & 15.759(060) & 14.624(057) \\
    58540.9  & 18.882(166) & 19.544(247) & 17.820(121) & 16.495(080) & 16.050(062) & 14.858(059) \\
    58543.2  & 19.075(200) & 19.505(263) & 17.891(134) & 16.590(087) & 16.086(065) & 14.888(062) \\

 \hline
\multicolumn{5}{l}{Note: Uncertainties, in units of 0.001 mag, are $1\sigma$.}\\
\label{swift}
\end{tabular}
\end{table*}

\begin{table*}
\centering
\caption{Log of Optical Spectroscopic Observations of SN\,2019np  \label{log} }
\begin{tabular}{ccccccccc}
\hline
MJD &Phase$^{a}$ & Range (\AA) & Resolution. (\AA)  & Telescope+Inst.\\
\hline
    58493.6  & -16.6  & 3477-8727 & 25    & LJT+YFOSC \\
    58493.7  & -16.5  & 3477-8727 & 25    & LJT+YFOSC \\
    58495.7  & -14.5  & 3838-8770 & 15    & XLT+BFOSC \\
    58496.1  & -14.1  & 3385-9639 & 18    & NOT+ALFOSC  \\
    58497.2  & -13.0  & 3385-9640 & 18    & NOT+ALFOSC  \\
    58498.2  & -12.0  & 3385-9606 & 18    & NOT+ALFOSC  \\
    58499.7  & -10.5  & 3487-8730 & 25    & LJT+YFOSC \\
    58503.2  & -7.0  & 3385-9639 & 18    & NOT+ALFOSC  \\
    58506.2  & -4.0  & 3386-9604 & 18    & NOT+ALFOSC  \\
    58506.9  & -3.3  & 3483-8727 & 25    & LJT+YFOSC \\
    58507.9  & -2.4  & 3833-8770 & 15    & XLT+BFOSC \\
    58509.2  & -1.1  & 3385-9640 & 18    & NOT+ALFOSC  \\
    58512.1  & 1.9   & 3385-9635 & 18    & NOT+ALFOSC  \\
    58518.7  & 8.5   & 3484-8727 & 25    & LJT+YFOSC \\
    58520.0  & 9.8   & 3385-9605 & 18    & NOT+ALFOSC  \\
    58521.7  & 11.5  & 3488-8727 & 25    & LJT+YFOSC \\
    58522.7  & 12.5  & 4347-8665 & 15    & XLT+BFOSC \\
    58524.7  & 14.5  & 3485-8728 & 25    & LJT+YFOSC \\
    58525.6  & 15.4  & 4348-8666 & 15    & XLT+BFOSC \\
    58526.9  & 16.7  & 3420-9255 & 24    & EKAR+AFOSC \\
    58528.7  & 18.5  & 3489-8727 & 25    & LJT+YFOSC \\
    58536.0  & 25.8  & 3385-9606 & 18    & NOT+ALFOSC  \\
    58539.9  & 29.7  & 3498-9256 & 24    & EKAR+AFOSC \\
    58543.0  & 32.8  & 3385-9253 & 24    & EKAR+AFOSC \\
    58548.0  & 37.8  & 3385-9242 & 24    & EKAR+AFOSC \\
    58552.7  & 42.5  & 3489-8729 & 25    & LJT+YFOSC \\
    58554.8  & 44.6  & 3841-8783 & 15    & XLT+BFOSC \\
    58557.1  & 46.9  & 3580-9658 & 18    & NOT+ALFOSC  \\
    58565.8  & 55.6  & 4373-8666 & 15    & XLT+BFOSC \\
    58570.6  & 60.4  & 4071-8769 & 15    & XLT+BFOSC \\
    58571.0  & 60.8  & 3389-9246 & 24    & EKAR+AFOSC \\
    58601.9  & 91.7  & 3403-9650 & 18    & NOT+ALFOSC  \\
    58608.6  & 98.4  & 4077-8772 & 15    & XLT+BFOSC \\
    58617.9  & 107.7  & 3402-9649 & 18    & NOT+ALFOSC  \\
    58637.9  & 127.7  & 3484-9644 & 18    & NOT+ALFOSC  \\
    58653.9  & 143.7  & 3534-9053 & 18    & NOT+ALFOSC  \\
    58809.2  & 298.9  & 3891-8826 & 18    & NOT+ALFOSC  \\
    58813.2  & 303.0  & 3813-7831 & 7     & GTC+OSIRIS \\
    58878.0  & 367.8  & 3833-7831 & 7     & GTC+OSIRIS \\
    
\hline
\multicolumn{5}{l}{$^a$ Days relative to the $B-$band maximum on MJD 58510.2.}\\
\end{tabular}
\end{table*}

\begin{table*}
\centering
\caption{Log of NIR Spectroscopic Observations of SN\,2019np  \label{lognir} }
\begin{tabular}{ccccccccc}
\hline
MJD &Phase$^{a}$ & Range (\AA) & R  & Telescope+Inst.\\
\hline
    58533.2 &23.0&7839-25345& 450 &FIRE\\
    58566.2 &56.0&6835-25343& 1200 &IRTF\\
    58584.2 &74.0&6844-25336& 1200 &IRTF\\
    58619.0 &108.8&7887-25276& 450 &FIRE\\
\hline
\multicolumn{5}{l}{$^a$ Days relative to the $B-$band maximum on MJD 58510.2.}\\
\end{tabular}
\end{table*}

\begin{landscape}
\begin{table}
\centering
\caption{Ground-based Optical Photometry of SN\,2019np}
\begin{tabular}{ccccccccccccc}
\hline
MJD & \textit{B} (mag) & \textit{V} (mag) & \textit{R} (mag) & \textit{I} (mag) & \textit{u} (mag) & \textit{g} (mag) & \textit{r} (mag) & \textit{i} (mag) & \textit{z} (mag) & data source  \\
\hline
    58492.5  & ...   & ...   & ...   & ...   & ...   & ...   & 18.177(131) & ...   & ...   & ZTF \\
    58492.7  & ...   & ...   & 17.401(161) & ...   & ...   & ...   & ...   & ...   & ...   & BITRAN-CCD \\
    58493.4  & ...   & ...   & ...   & ...   & ...   & ...   & 17.001(033) & ...   & ...   & ZTF \\
    58493.7  & 16.977(043) & 16.701(078) & ...   & ...   & ...   & 16.820(060) & 16.738(080) & 16.950(100) & ...   & LJT \\
    58494.5  & ...   & ...   & ...   & ...   & ...   & 16.356(036) & ...   & ...   & ...   & ZTF \\
    58494.9  & ...   & ...   & 16.263(011) & 16.000(022) & ...   & ...   & ...   & ...   & ...   & MEIA3 \\
    58495.2  & 16.034(027) & 15.890(039) & 16.229(080) & 15.796(036) & ...   & ...   & ...   & ...   & ...   & MEIA3 \\
    58495.9  & ...   & 15.735(013) & 15.797(027) & ...   & ...   & ...   & ...   & ...   & ...   & MEIA3 \\
    58495.9  & 15.999(021) & 15.816(011) & ...   & ...   & ...   & 15.781(010) & 15.904(010) & 16.157(013) & ...   & TNT \\
    58496.2  & 15.668(023) & 15.444(025) & 15.545(044) & 15.308(020) & ...   & ...   & ...   & ...   & ...   & MEIA3 \\
    58496.7  & 15.671(038) & 15.524(014) & ...   & ...   & ...   & 15.482(016) & 15.596(018) & 15.800(012) & ...   & TNT \\
    58497.2  & ...   & 15.258(016) & ...   & ...   & ...   & ...   & ...   & ...   & ...   & MEI3A \\
    58497.7  & 15.208(016) & 15.155(007) & ...   & ...   & ...   & 15.065(005) & 15.186(004) & 15.378(005) & ...   & TNT \\
    58498.0  & 14.888(028) & 14.977(022) & 15.083(037) & 14.776(024) & ...   & ...   & ...   & ...   & ...   & MEIAA \\
    58498.1  & 14.812(019) & 14.922(022) & 15.060(053) & 14.682(028) & ...   & ...   & ...   & ...   & ...   & MEIA3 \\
    58498.8  & 14.833(018) & 14.802(006) & ...   & ...   & ...   & 14.717(007) & 14.838(005) & 15.020(007) & ...   & TNT \\
    58498.9  & 14.600(042) & 14.785(044) & 14.929(111) & 14.547(043) & ...   & ...   & ...   & ...   & ...   & MEIA3 \\
    58499.0  & 14.409(051) & 14.610(040) & 14.855(111) & 14.518(042) & ...   & ...   & ...   & ...   & ...   & MEIA3 \\
    58499.7  & 14.444(051) & 14.544(011) & ...   & ...   & ...   & 14.394(029) & 14.644(017) & 15.069(133) & ...   & LJT \\
    58499.8  & 14.529(014) & 14.514(006) & ...   & ...   & ...   & 14.356(007) & 14.555(005) & 14.759(007) & ...   & TNT \\
    58500.0  & ...   & ...   & ...   & 14.159(029) & ...   & ...   & ...   & ...   & ...   & MEIA3 \\
    58500.8  & 14.279(013) & 14.299(006) & ...   & ...   & ...   & 14.164(006) & 14.321(006) & 14.526(005) & ...   & TNT \\
    58501.0  & 13.962(026) & 14.117(028) & 14.277(056) & 14.009(024) & ...   & ...   & ...   & ...   & ...   & MEIA3 \\
    58502.0  & 13.695(027) & ...   & 14.046(024) & 13.846(028) & ...   & ...   & ...   & ...   & ...   & MEIA3 \\
    58502.4  & ...   & ...   & ...   & ...   & ...   & 13.862(020) & ...   & ...   & ...   & ZTF \\
    58502.5  & ...   & ...   & ...   & ...   & ...   & ...   & 13.919(027) & ...   & ...   & ZTF \\
    58503.8  & 13.831(026) & 13.859(013) & ...   & ...   & ...   & 13.710(014) & 13.847(008) & 14.076(008) & ...   & TNT \\
    58504.0  & ...   & 13.726(023) & ...   & ...   & ...   & ...   & ...   & ...   & ...   & MEIA3 \\
    58504.1  & 13.491(024) & 13.723(030) & 13.789(057) & 13.612(029) & ...   & ...   & ...   & ...   & ...   & MEIA3 \\
    58504.8  & 13.693(026) & 13.714(016) & ...   & ...   & ...   & 13.593(005) & 13.727(006) & 13.998(006) & ...   & TNT \\
    58504.9  & ...   & 13.627(012) & ...   & 13.511(019) & ...   & ...   & ...   & ...   & ...   & MEIA3 \\
    58505.0  & 13.403(016) & 13.646(014) & 13.539(022) & ...   & ...   & ...   & ...   & ...   & ...   & MEIA3 \\
    58505.1  & 13.395(015) & 13.648(017) & 13.604(038) & 13.494(020) & ...   & ...   & ...   & ...   & ...   & MEIA3 \\
    58506.0  & 13.321(019) & 13.684(029) & 13.493(059) & 13.463(024) & ...   & ...   & ...   & ...   & ...   & MEIA3 \\
    58506.9  & 13.19(0130) & 13.683(035) & ...   & ...   & ...   & 13.505(064) & 14.049(085) & 13.944(060) & ...   & LJT \\
    58507.1 & 13.407(004) & 13.534(003) & 13.328(023) & 13.491(004) & ...   & ...   & ...   & ...   & ...   & ALFOSC\_FASU \\
    58507.5  & ...   & ...   & ...   & ...   & ...   & 13.412(016) & ...   & ...   & ...   & ZTF \\
    58508.9  & 13.537(022) & 13.473(012) & ...   & ...   & ...   & 13.387(009) & 13.570(008) & 13.985(011) & ...   & TNT \\
    58510.3  & ...   & ...   & ...   & ...   & ...   & ...   & 13.468(026) & ...   & ...   & ZTF \\
    58511.4  & ...   & ...   & ...   & ...   & ...   & ...   & 13.457(031) & ...   & ...   & ZTF \\
    58511.7  & 13.518(015) & 13.448(006) & ...   & ...   & ...   & 13.383(009) & 13.540(010) & 14.120(007) & ...   & TNT \\
    58512.8  & 13.520(015) & 13.470(006) & ...   & ...   & ...   & 13.416(005) & 13.556(003) & 14.159(006) & ...   & TNT \\
    58513.4  & ...   & ...   & ...   & ...   & ...   & ...   & 13.491(026) & ...   & ...   & ZTF \\
    58514.8  & 13.633(015) & 13.508(006) & ...   & ...   & ...   & 13.478(008) & 13.584(006) & 14.227(007) & ...   & TNT \\
    58515.8  & 13.667(013) & 13.537(006) & ...   & ...   & ...   & 13.521(004) & 13.628(003) & 14.276(006) & ...   & TNT \\

\hline
\multicolumn{5}{l}{Note: Uncertainties, in units of 0.001 mag, are $1\sigma$.}\\
 \end{tabular}%
  \label{tab:optical}%
\end{table}%
\end{landscape}

\begin{landscape}
\begin{table}
 \contcaption{A table continued from the previous one.}
 \label{tab:continued}
\begin{tabular}{ccccccccccccc}
\hline
MJD & \textit{B} (mag) & \textit{V} (mag) & \textit{R} (mag) & \textit{I} (mag) & \textit{u} (mag) & \textit{g} (mag) & \textit{r} (mag) & \textit{i} (mag) & \textit{z} (mag) & data source \\
\hline
    58517.7  & 13.851(016) & 13.631(008) & ...   & ...   & ...   & 13.595(009) & ...   & 14.446(006) & ...   & TNT \\
    58518.2  & 13.552(015) & 13.576(011) & 13.630(020) & ...   & ...   & ...   & ...   & ...   & ...   & MEIA3 \\
    58518.7  & 13.852(018) & 13.531(043) & ...   & ...   & ...   & 13.844(034) & 13.769(026) & 14.538(033) & ...   & LJT \\
    58519.0  & 13.625(017) & 13.572(016) & 13.698(041) & 13.963(036) & ...   & ...   & ...   & ...   & ...   & MEIA3 \\
    58520.0  & 13.791(012) & 13.637(021) & 13.751(004) & 14.160(004) & ...   & ...   & ...   & ...   & ...   & ALFOSC\_FASU \\
    58520.1  & 13.722(017) & 13.679(018) & 13.862(044) & 14.041(028) & ...   & ...   & ...   & ...   & ...   & MEIA3 \\
    58520.8  & 13.930(026) & 13.717(028) & 14.225(077) & 14.131(046) & ...   & ...   & ...   & ...   & ...   & MEIA3 \\
    58521.7  & 14.030(026) & 13.674(024) & ...   & ...   & ...   & 13.963(033) & 13.912(025) & 14.758(024) & ...   & LJT \\
    58521.8  & 13.870(017) & 13.746(019) & 14.070(048) & 14.129(020) & ...   & ...   & ...   & ...   & ...   & MEIA3 \\
    58522.4  & ...   & ...   & ...   & ...   & ...   & 13.895(029) & ...   & ...   & ...   & ZTF \\
    58524.0  & 14.282(036) & 13.773(051) & 14.382(040) & 14.262(040) & ...   & ...   & ...   & ...   & ...   & MEIA3 \\
    58524.7  & 14.329(052) & 13.874(038) & ...   & ...   & ...   & 14.102(026) & 14.120(015) & 14.881(027) & ...   & LJT \\
    58524.9  & ...   & 13.958(020) & ...   & ...   & ...   & ...   & ...   & ...   & ...   & MEIA3 \\
    58525.7  & 14.627(020) & 14.059(007) & ...   & ...   & ...   & 14.156(014) & 14.204(035) & 14.924(006) & ...   & TNT \\
    58526.0  & ...   & ...   & 14.083(014) & 14.168(022) & ...   & ...   & ...   & ...   & ...   & MEIA3 \\
    58526.4  & ...   & ...   & ...   & ...   & ...   & 14.257(020) & ...   & ...   & ...   & ZTF \\
    58526.9  & 14.461(067) & 14.072(021) & ...   & ...   & 15.519(024) & 14.321(018) & 14.208(014) & 14.791(012) & 14.385(014) & AFOSC \\
    58527.1  & 14.371(025) & 14.071(027) & 14.290(095) & 14.141(028) & ...   & ...   & ...   & ...   & ...   & MEIA3 \\
    58527.9  & ...   & ...   & ...   & ...   & 15.765(055) & 14.272(028) & 14.114(024) & 14.852(054) & ...   & Moravian \\
    58528.0  & 14.621(028) & 14.219(039) & ...   & ...   & ...   & ...   & ...   & ...   & ...   & Moravian \\
    58528.6  & 14.873(042) & 14.165(030) & ...   & ...   & ...   & 14.369(025) & 14.206(021) & 14.812(023) & ...   & LJT \\
    58528.9  & 14.626(025) & 14.311(040) & ...   & ...   & 15.982(055) & 14.464(048) & 14.218(027) & 14.763(048) & ...   & Moravian \\
    58529.9  & 14.852(039) & 14.197(030) & 14.350(079) & 14.177(055) & ...   & ...   & ...   & ...   & ...   & MEIA3 \\
    58530.9  & 15.238(018) & 14.355(007) & ...   & ...   & ...   & 14.642(005) & 14.350(04) & 14.811(006) & ...   & TNT \\
    58530.9  & 15.039(044) & 14.344(035) & 14.251(066) & 14.179(053) & ...   & ...   & ...   & ...   & ...   & MEIA3 \\
    58533.0  & ...   & 14.354(064) & 14.223(067) & 14.120(033) & ...   & ...   & ...   & ...   & ...   & MEIA3 \\
    58533.8  & 15.584(047) & 14.484(027) & ...   & ...   & ...   & 14.941(022) & ...   & ...   & ...   & TNT \\
    58534.4  & ...   & ...   & ...   & ...   & ...   & ...   & 14.378(019) & ...   & ...   & ZTF \\
    58534.7  & 15.618(016) & 14.562(007) & ...   & ...   & ...   & 14.996(005) & 14.430(006) & 14.726(007) & ...   & TNT \\
    58535.7  & 15.718(015) & 14.599(007) & ...   & ...   & ...   & 15.053(006) & 14.433(005) & 14.720(007) & ...   & TNT \\
    58536.8  & 15.527(031) & 14.607(032) & ...   & ...   & ...   & ...   & ...   & ...   & ...   & MEIA3 \\
    58538.3  & ...   & ...   & ...   & ...   & ...   & 15.278(031) & 14.500(035) & ...   & ...   & ZTF \\
    58538.7  & 15.996(017) & 14.762(007) & ...   & ...   & ...   & 15.313(005) & 14.558(008) & 14.687(007) & ...   & TNT \\
    58538.9  & 15.567(016) & 14.767(021) & 14.362(115) & 14.039(023) & ...   & ...   & ...   & ...   & ...   & MEIA3 \\
    58540.0  & 15.892(027) & 14.801(034) & 14.329(045) & 14.015(022) & ...   & ...   & ...   & ...   & ...   & MEIA3 \\
    58540.0  & ...   & ...   & ...   & ...   & 16.891(073) & 15.405(025) & 14.521(015) & 14.604(018) & 14.350(016) & AFOSC \\
    58540.0  & 15.725(057) & 14.753(010) & ...   & ...   & ...   & ...   & ...   & ...   & ...   & AFOSC \\
    58540.7  & 15.929(087) & 15.035(070) & ...   & ...   & ...   & 15.483(056) & 14.744(072) & 14.950(095) & ...   & LJT \\
    58541.3  & ...   & ...   & ...   & ...   & ...   & 15.505(031) & 14.627(037) & ...   & ...   & ZTF \\
    58541.6  & 16.187(016) & 14.946(007) & ...   & ...   & ...   & 15.526(006) & 14.705(007) & 14.783(007) & ...   & TNT \\
    58542.0  & 15.808(035) & 14.931(020) & 14.559(059) & 14.131(028) & ...   & ...   & ...   & ...   & ...   & MEIA3 \\
    58542.9  & 16.053(040) & 14.905(027) & 14.720(074) & 14.235(042) & ...   & ...   & ...   & ...   & ...   & MEIA3 \\
    58543.0  & ...   & ...   & ...   & ...   & 17.444(024) & 15.562(036) & 14.628(029) & 14.631(034) & 14.380(037) & AFOSC \\
    58543.0  & 15.926(034) & 14.895(027) & ...   & ...   & 17.197(043) & 15.566(019) & 14.744(024) & 14.668(030) & ...   & Moravian \\
    58544.0  & 15.913(042) & 14.963(032) & ...   & ...   & 17.238(043) & 15.532(026) & 14.756(027) & 14.702(039) & ...   & Moravian \\
    58544.7  & 16.327(016) & 15.097(007) & ...   & ...   & ...   & 15.660(010) & 14.908(006) & 14.979(007) & ...   & TNT \\
\hline
\multicolumn{5}{l}{Note: Uncertainties, in units of 0.001 mag, are $1\sigma$.}\\
 \end{tabular}%
  \label{tab:addlabel}%
\end{table}%
\end{landscape}

\begin{landscape}
\begin{table}
 \contcaption{A table continued from the previous one.}
 \label{tab:continued}
\begin{tabular}{ccccccccccccc}
\hline
MJD & \textit{B} (mag) & \textit{V} (mag) & \textit{R} (mag) & \textit{I} (mag) & \textit{u} (mag) & \textit{g} (mag) & \textit{r} (mag) & \textit{i} (mag) & \textit{z} (mag) & data source \\
\hline
    58545.1  & 15.984(023) & 15.023(031) & 14.719(060) & ...   & ...   & ...   & ...   & ...   & ...   & MEIA3 \\
    58546.0  & 15.976(102) & 15.073(030) & ...   & ...   & 17.276(043) & 15.696(017) & 14.880(021) & 14.834(018) & 14.551(031) & AFOSC \\
    58547.1  & 16.152(037) & 15.080(026) & ...   & ...   & 17.307(117) & 15.795(066) & 14.973(026) & 15.012(031) & ...   & Moravian \\
    58548.0  & 16.171(025) & 15.123(033) & ...   & ...   & 17.455(060) & 15.795(031) & 15.018(033) & 15.055(035) & ...   & Moravian \\
    58548.8  & 16.538(020) & 15.328(008) & ...   & ...   & ...   & 15.893(008) & 15.165(009) & 15.244(007) & ...   & TNT \\
    58550.9  & 16.321(032) & 15.361(027) & 15.254(083) & 14.707(040) & ...   & ...   & ...   & ...   & ...   & MEIA3 \\
    58552.7  & 16.478(064) & 15.511(038) & ...   & ...   & ...   & 15.975(034) & 15.407(130) & 15.464(133) & ...   & LJT \\
    58553.6  & 16.671(018) & 15.503(007) & ...   & ...   & ...   & 16.045(004) & 15.392(003) & 15.504(009) & ...   & TNT \\
    58553.9  & 16.383(032) & 15.413(022) & 15.308(079) & 14.833(038) & ...   & ...   & ...   & ...   & ...   & MEIA3 \\
    58556.9  & 16.449(026) & 15.487(010) & 15.240(050) & 14.897(018) & ...   & ...   & ...   & ...   & ...   & MEIA3 \\
    58557.1  & 16.400(005) & 15.500(004) & ...   & ...   & ...   & ...   & 15.395(015) & 15.443(041) & ...   & ALFOSC\_FASU \\
    58558.2  & ...   & ...   & ...   & ...   & ...   & ...   & 15.522(025) & ...   & ...   & ZTF \\
    58559.2  & ...   & ...   & ...   & ...   & ...   & ...   & 15.539(034) & ...   & ...   & ZTF \\
    58559.7  & 16.622(089) & 15.661(045) & ...   & ...   & ...   & 16.172(048) & 15.575(032) & 15.781(064) & ...   & TNT \\
    58559.9  & ...   & 15.594(022) & 15.335(051) & ...   & ...   & ...   & ...   & ...   & ...   & MEIA3 \\
    58563.0  & 16.527(051) & 15.549(044) & 15.517(131) & 15.259(053) & ...   & ...   & ...   & ...   & ...   & MEIA3 \\
    58563.5  & 16.873(034) & 15.803(008) & ...   & ...   & ...   & 16.279(015) & 15.757(011) & 15.987(012) & ...   & TNT \\
    58566.0  & 16.485(041) & 15.812(042) & 15.665(052) & 15.446(043) & ...   & ...   & ...   & ...   & ...   & MEIA3 \\
    58566.6  & 16.915(016) & 15.895(007) & ...   & ...   & ...   & 16.292(006) & 15.855(007) & 16.060(011) & ...   & TNT \\
    58569.0  & 16.670(014) & 15.776(011) & 15.676(026) & 15.399(023) & ...   & ...   & ...   & ...   & ...   & MEIA3 \\
    58572.0  & 16.615(021) & 15.815(021) & 15.797(053) & 15.549(032) & ...   & ...   & ...   & ...   & ...   & MEIA3 \\
    58572.2  & ...   & ...   & ...   & ...   & ...   & ...   & 15.989(027) & ...   & ...   & ZTF \\
    58572.5  & 16.972(021) & 16.063(009) & ...   & ...   & ...   & 16.444(010) & 16.067(012) & 16.294(014) & ...   & TNT \\
    58574.2  & ...   & ...   & ...   & ...   & ...   & ...   & 16.099(057) & ...   & ...   & ZTF \\
    58575.2  & ...   & ...   & ...   & ...   & ...   & ...   & 16.095(062) & ...   & ...   & ZTF \\
    58577.6  & 17.075(019) & 16.179(007) & ...   & ...   & ...   & 16.520(007) & 16.241(009) & 16.490(016) & ...   & TNT \\
    58578.0  & ...   & 16.250(042) & 15.985(085) & 15.857(080) & ...   & ...   & ...   & ...   & ...   & MEIA3 \\
    58580.6  & …     & 16.250(018) & ...   & ...   & ...   & 16.582(009) & 16.338(009) & 16.620(009) & ...   & TNT \\
    58581.1  & 16.812(029) & 16.135(043) & 16.095(109) & ...   & ...   & ...   & ...   & ...   & ...   & MEIA3 \\
    58581.2  & ...   & ...   & ...   & ...   & ...   & ...   & 16.271(037) & ...   & ...   & ZTF \\
    58582.2  & ...   & ...   & ...   & ...   & ...   & ...   & 16.320(030) & ...   & ...   & ZTF \\
    58584.0  & 16.836(021) & 16.175(013) & 16.145(073) & 15.977(019) & ...   & ...   & ...   & ...   & ...   & MEIA3 \\
    58584.2  & ...   & ...   & ...   & ...   & ...   & ...   & 16.363(032) & ...   & ...   & ZTF \\
    58587.3  & ...   & ...   & ...   & ...   & ...   & ...   & 16.455(024) & ...   & ...   & ZTF \\
    58589.6  & 17.24(126) & 16.450(068) & ...   & ...   & ...   & 16.666(073) & 16.687(072) & 16.949(022) & ...   & TNT \\
    58590.0  & ...   & 16.335(018) & ...   & ...   & ...   & ...   & ...   & ...   & ...   & MEIA3 \\
    58590.7  & 17.411(115) & 16.414(056) & ...   & ...   & ...   & 16.795(092) & ...   & 16.921(026) & ...   & TNT \\
    58591.4  & ...   & ...   & ...   & ...   & ...   & ...   & 16.603(028) & ...   & ...   & ZTF \\
    58592.7  & 17.231(122) & 16.601(042) & ...   & ...   & ...   & 16.811(068) & 16.819(099) & 16.883(060) & ...   & LJT \\
    58593.3  & ...   & ...   & ...   & ...   & ...   & ...   & 16.650(041) & ...   & ...   & ZTF \\
    58594.2  & ...   & ...   & ...   & ...   & ...   & ...   & 16.673(042) & ...   & ...   & ZTF \\
    58594.7  & 17.242(082) & 16.614(048) & ...   & ...   & ...   & 16.726(039) & 16.701(045) & 16.971(075) & ...   & TNT \\
    58597.2  & ...   & ...   & ...   & ...   & ...   & ...   & 16.764(044) & ...   & ...   & ZTF \\
    58597.4  & ...   & ...   & ...   & ...   & ...   & ...   & 16.772(040) & ...   & ...   & ZTF \\
    58601.3  & ...   & ...   & ...   & ...   & ...   & ...   & 16.891(037) & ...   & ...   & ZTF \\
\hline
\multicolumn{5}{l}{Note: Uncertainties, in units of 0.001 mag, are $1\sigma$.}\\
 \end{tabular}%
  \label{tab:addlabel}%
\end{table}%
\end{landscape}

\begin{landscape}
\begin{table}
 \contcaption{A table continued from the previous one.}
 \label{tab:continued}
\begin{tabular}{ccccccccccccc}
\hline
MJD & \textit{B} (mag) & \textit{V} (mag) & \textit{R} (mag) & \textit{I} (mag) & \textit{u} (mag) & \textit{g} (mag) & \textit{r} (mag) & \textit{i} (mag) & \textit{z} (mag) & data source \\
\hline
    58603.6  & 17.433(016) & 16.793(008) & ...   & ...   & ...   & 16.929(009) & 17.017(007) & 17.391(026) & ...   & TNT \\
    58605.5  & 17.392(024) & 16.835(010) & ...   & ...   & ...   & 16.958(014) & 17.118(008) & 17.423(030) & ...   & TNT \\
    58606.7  & 17.368(041) & 16.821(027) & ...   & ...   & ...   & ...   & 17.026(032) & 17.282(041) & ...   & LJT \\
    58608.2  & ...   & ...   & ...   & ...   & ...   & ...   & 17.094(044) & ...   & ...   & ZTF \\
    58608.6  & 17.426(022) & 16.946(008) & ...   & ...   & ...   & 17.031(009) & 17.202(009) & 17.519(033) & ...   & TNT \\
    58612.2  & ...   & ...   & ...   & ...   & ...   & ...   & 17.201(045) & ...   & ...   & ZTF \\
    58617.2  & ...   & ...   & ...   & ...   & ...   & ...   & 17.327(055) & ...   & ...   & ZTF \\
    58618.3  & ...   & ...   & ...   & ...   & ...   & ...   & 17.373(042) & ...   & ...   & ZTF \\
    58619.7  & 17.598(211) & 17.098(052) & ...   & ...   & ...   & 17.281(105) & 17.334(047) & 17.888(046) & ...   & LJT \\
    58623.6  & 17.671(031) & 17.360(010) & ...   & ...   & ...   & 17.333(017) & 17.695(009) & 18.107(052) & ...   & TNT \\
    58630.6  & 17.741(028) & 17.475(011) & ...   & ...   & ...   & 17.400(017) & 17.861(018) & 18.310(066) & ...   & TNT \\
    58633.6  & 17.789(034) & 17.535(012) & ...   & ...   & ...   & 17.492(022) & 17.931(034) & 18.332(057) & ...   & TNT \\
    58634.2  & ...   & ...   & ...   & ...   & ...   & ...   & 17.825(067) & ...   & ...   & ZTF \\
    58637.2  & ...   & ...   & ...   & ...   & ...   & ...   & 17.908(050) & ...   & ...   & ZTF \\
    58637.9  & 17.704(007) & 17.469(016) & ...   & ...   & ...   & ...   & 17.831(017) & 18.188(024) & ...   & ALFOSC\_FASU \\
    58640.2  & ...   & ...   & ...   & ...   & ...   & ...   & 17.971(065) & ...   & ...   & ZTF \\
    58645.5  & 17.791(131) & 17.568(016) & ...   & ...   & ...   & 17.544(018) & 18.310(015) & 18.545(086) & ...   & TNT \\
    58647.2  & ...   & ...   & ...   & ...   & ...   & ...   & 18.145(091) & ...   & ...   & ZTF \\
    58648.6  & 18.017(072) & 17.899(018) & ...   & ...   & ...   & 17.610(016) & 18.314(025) & 18.746(103) & ...   & TNT \\
    58650.2  & ...   & ...   & ...   & ...   & ...   & ...   & 18.194(089) & ...   & ...   & ZTF \\
    58653.9  & 17.957(031) & 17.767(020) & ...   & ...   & ...   & ...   & 18.270(014) & 18.514(017) & ...   & ALFOSC\_FASU \\
    58816.8  & ...   & ...   & ...   & ...   & ...   & 19.115(073) & 19.045(104) & 19.008(080) & ...   & LJT \\
    58929.7  & ...   & ...   & ...   & ...   & ...   & 20.307(123) & 19.816(085) & 19.525(086) & ...   & LJT \\
\hline
\multicolumn{5}{l}{Note: Uncertainties, in units of 0.001 mag, are $1\sigma$.}\\
 \end{tabular}%
  \label{tab:addlabel}%
\end{table}%
\end{landscape}

\bsp	
\label{lastpage}
\end{document}